\long\def\symbolfootnote[#1]#2{\begingroup%
\def\thefootnote{\fnsymbol{footnote}}\footnote[#1]{#2}\endgroup} 
\long\def\symbolfootnote[#1]#2{\begingroup%
\def\thefootnote{\fnsymbol{footnote}}\footnote[#1]{#2}\endgroup}

\def\kms{\text{km} \ \text{s}^{-1}}
\def\ms{\text{m} \ \text{s}^{-1}}

\def\nbody{$N$-body\ }
\def\reff{r_{\rm eff}}
\def\feh{\rm{[Fe/H]}}
\def\MV{M_V}
\def\MVi{M_{V_i}}
\def\ml{M/L_V}
\def\Mi{M_{\rm i}}
\def\pc{{\rm pc}}
\def\kpc{{\rm kpc}}

\def\Myr{{\rm Myr}}
\def\Gyr{{\rm Gyr}}
\def\dr{{\rm d}}
\def\tdis{t_{\rm dis}}
\def\trh{t_{\rm rh}}
\def\trhi{t_{\rm rh,0}}
\def\rh{r_{\rm h}}
\def\rJ{r_{\rm J}}
\def\rj{r_{\rm J}}
\def\age{Age}
\def\msun{{\rm M_\odot}}
\def\Msun{{\rm M_\odot}}
\def\lsun{{\rm L_\odot}}
\def\Lsun{{\rm L_\odot}}
\def\mlsun{{\rm \msun/\lsun}}
\def\rhoh{\rho_{\rm h}}
\def\Nh{N_{\rm h}}
\def\RG{R_{\rm G}}
\def\Rp{R_{\rm p}}
\def\Ra{R_{\rm a}}
\def\RC{R_{\rm C}}
\def\VC{V_{\rm C}}
\def\nP{n_{\rm P}}
\def\nsel{N_{\rm GC}}
\def\nBG{n_{\rm BG}}
\def\sigj{\sigma_j}
\def\trilegal{{\sc trilegal}}
\def\nbound{N_{\rm bound}}

\def\Ms{M_{\rm s}}
\def\Rs{R_{\rm s}}
\def\Mb{M_{\rm b}}
\def\ab{a_{\rm b}}
\def\bb{b_{\rm b}}
\def\Md{M_{\rm d}}
\def\ad{a_{\rm d}}
\def\bd{b_{\rm d}}
\def\Mh{M_{\rm h}}
\def\NWD{N_{\rm WD}}
\def\N*{N_{\rm *}}

\def\ft{f_{\rm t}}

\documentclass[useAMS,usenatbib]{mnras}
\usepackage{amssymb,latexsym,graphicx,natbib,eufrak,amsmath}

\usepackage{color}

\usepackage{setspace}

\usepackage[english]{babel}
\usepackage{nicefrac,soul}

\usepackage{makecell}
\usepackage{caption}

\title[]
  {
The contribution of dissolving star clusters to the population of ultra-faint objects in the outer halo of the Milky Way
  }
\author[F. Contenta et al.]
  {Filippo Contenta, Mark Gieles, Eduardo Balbinot, Michelle L. M. Collins\\
  Department of Physics, University of Surrey, Guildford GU2 7XH, UK}
\pagerange{\pageref{firstpage}--\pageref{lastpage}} \pubyear{2016}

\def\LaTeX{L\kern-.36em\raise.3ex\hbox{a}\kern-.15em
    T\kern-.1667em\lower.7ex\hbox{E}\kern-.125emX}

\begin{document}
\maketitle
\begin{abstract}
{In the last decade, several ultra faint objects (UFOs, $\MV\gtrsim-3.5$) have been discovered in the outer halo of the Milky Way. For some of these objects it is not clear whether they are star clusters or (ultra-faint) dwarf galaxies. In this work we quantify the contribution of star clusters to the population of UFOs. We extrapolated the mass and Galactocentric radius distribution of the globular clusters using a population model, finding that the Milky Way contains about $3.3^{+7.3}_{-1.6}$ star clusters with \mbox{$\MV\gtrsim-3.5$} and  Galactocentric radius $\geq20\,\kpc$. To understand whether dissolving clusters can appear as UFOs, we run a suite of direct \nbody models, varying the orbit, the Galactic potential, the binary fraction and the black hole (BH) natal kick velocities. In the analyses, we consider observational biases such as: luminosity limit, field stars, and line-of-sight projection. We find that star clusters contribute to both the compact and the extended population of UFOs: clusters without BHs appear compact with radii $\sim5\,\pc$, while clusters that retain their BHs after formation have radii $\gtrsim20\,\pc$. The properties of the extended clusters are remarkably similar to those of dwarf galaxies: high inferred mass-to-light ratios due to binaries; binary properties mildly affected by dynamical evolution; no observable mass segregation; and flattened stellar mass function. We conclude that the slope of the stellar mass function as a function of Galactocentric radius and the presence/absence of cold streams can discriminate between DM free and DM dominated UFOs.}
\end{abstract}

\section{Introduction}\label{UFO_intro}
\begin{figure*} 
\center
\includegraphics[width=\textwidth]{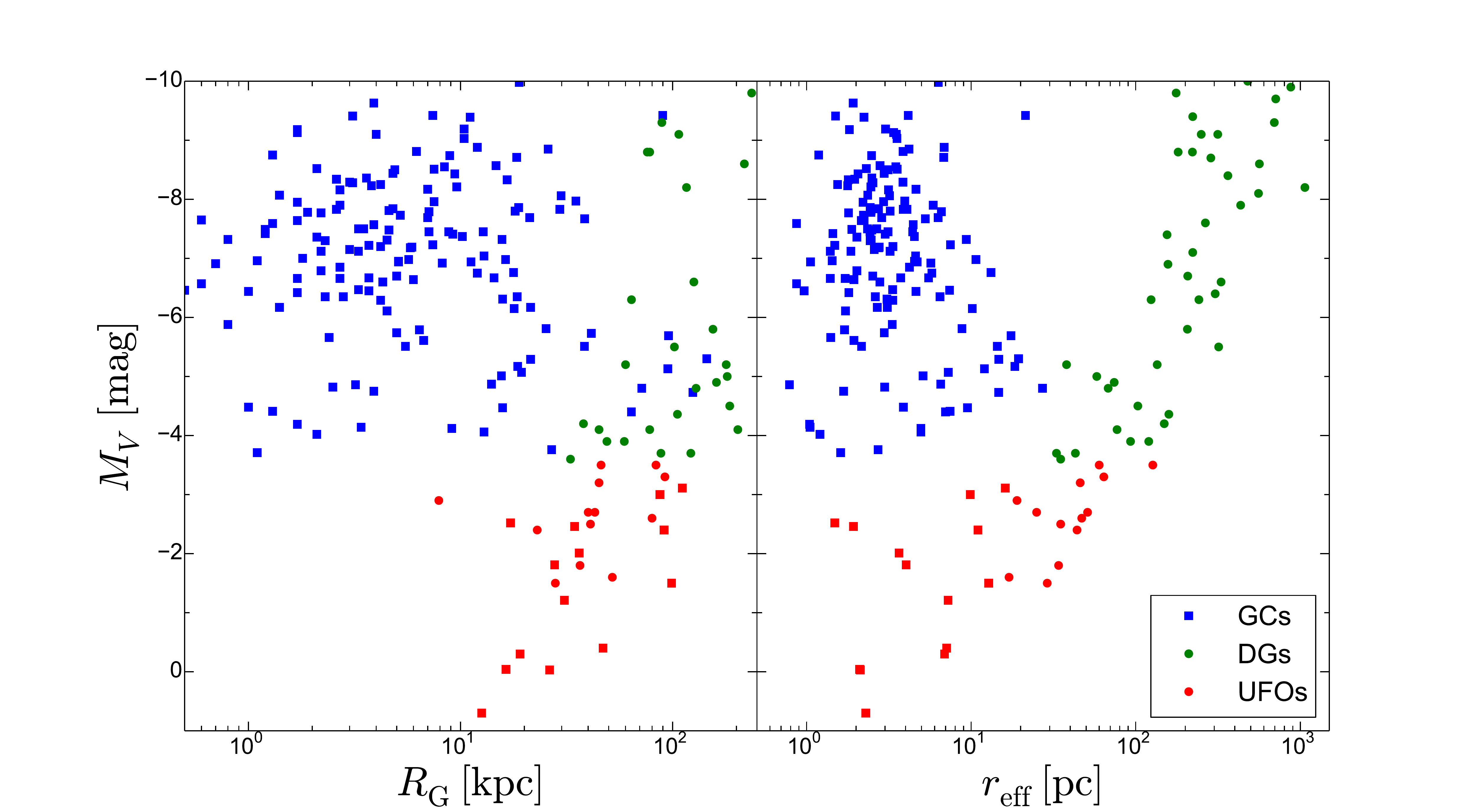}
\caption{
Distribution of Milky Way satellites in the Galactocentric distance-magnitude
space (left) and the size-magnitude space (right). GCs are shown as blue
squares, DGs are shown as green circles and the faint stellar systems, whose
nature has been topic of debate in literature, are in red $(\MV\gtrsim-3.5)$. The data on the GCs
were taken from \citet{2010arXiv1012.3224H}, these on the DGs from
\citet{2012AJ....144....4M} and the last satellites discovered were included. The recently discovered GCs (blue and red square) are: Segue 3, Mu{\~n}oz 1, Balbinot 1, Laevens 1/Crater, Laevens 3, Kim 1, Kim 2, Eridanus III, DES 1, Kim 3. While the recently discovered DGs (green and red circle) are: Hydra II, Laevens 2, Pegasus III, Ret II, Eridanus II, Tucana II, Horologium I, \mbox{Pictoris I}, Phoenix II, Draco II, Sagittarius II, Horologium II, Grus II, Tucana III, Columba I, Tucana IV, Reticulum III, \mbox{Tucana V}, Crater 2, Acquarius 2, Pictor II
[\citet{2011AJ....142...88F,2012ApJ...753L..15M,2013ApJ...767..101B,2014ApJ...786L...3L,
2014MNRAS.441.2124B,2014AJ....148...19P,
2015ApJ...799...73K,2015ApJ...802L..18L, 2015ApJ...813...44L, 
2015ApJ...804L...5M,2015ApJ...803...63K,2015ApJ...804L..44K,
2015ApJ...807...50B,2015ApJ...805..130K,2015arXiv150802381L,2015ApJ...813..109D,
2016ApJ...820..119K,2016MNRAS.459.2370T,2016MNRAS.463..712T,2016arXiv160902148D}].
\label{fig:lumVSradius}
}
\end{figure*}

The Milky Way halo contains numerous satellite stellar systems with a broad
range of luminosities. These stellar systems and their composition contain
valuable information about the formation of the Milky Way galaxy (e.g.
\citealt{1993ARA&A..31..575M, 2009ARA&A..47..371T, 2013NewAR..57..100B}).  Up to
a decade ago, there was a clear separation between dwarf galaxies (DGs) and
globular clusters (GCs).  In a diagram of absolute $V$-band magnitude ($\MV$)
vs. half-light radius ($\reff$, see Fig.~\ref{fig:lumVSradius}),  GCs (blue
squares) and DGs (green circles) with bright luminosities ($\MV \lesssim -3.5$)
are separated in size \citep{2007ApJ...663..948G}.  On the one hand, DGs are  large
($\reff\gtrsim30$\,pc), whereas GCs are compact  ($\reff\lesssim10$\,pc). In
addition, stars within DGs display a range of metallicities
($-3\lesssim\feh\lesssim-1.5$, fig. 12 in \citealt{2012AJ....144....4M}) and
their kinematics imply a high mass-to-light ratio, $10\lesssim\ml\lesssim1000$
(fig. 11 in \citealt{2012AJ....144....4M}), which is  usually explained by a
non-baryonic dark matter component
\citep{1998ARA&A..36..435M,2007ApJ...663..948G,2013pss5.book.1039W}.  Except for
a few exceptions, such as $\omega$ Cen \citep{1967RGOB..128..255D,1975ApJ...201L..71F,1978ApJ...225..148B} and M54 \citep{1995AJ....109.1086S}, GCs have no spread in iron abundance ($\feh$), however they do
display light-element anomalies \citep{2004ARA&A..42..385G} which are not seen
in DGs. Moreover, the internal kinematics of GCs can be explained by a single old stellar
populations with a `normal' initial stellar mass function (IMF), without the
need for dark matter \citep{2005ApJS..161..304M,2010ApJ...718..105D,2015MNRAS.448L..94S}.
Therefore, DGs and GCs were considered to be two totally different classes of
stellar systems.

Recently, thanks to the Sloan Digital Sky Survey
\citep[SDSS,][]{2000AJ....120.1579Y}, the Panoramic Survey Telescope And Rapid Response
System \citep[Pan-STARRS,][]{2015ApJ...802L..18L}, and the Dark Energy Survey \citep[DES,][]{2015ApJ...807...50B,2015ApJ...805..130K,2015ApJ...813..109D}, several ultra faint objects (UFOs,
$\MV\gtrsim-3.5$) have been discovered in the outer halo of the Milky Way (MW). For some MW satellites it is still debated whether they are GCs or ultra-faint DGs
\citep{2005AJ....129.2692W,2007ApJ...654..897B,2016MNRAS.458L..59M}. As shown in
Fig.~\ref{fig:lumVSradius}, at lower luminosities (objects marked in red) the
two populations (GCs and DGs) overlap at half-light radii of about $20$ to
$30\,\pc$ (hereafter, we refer to UFO with $\reff\ge 20\,$pc as extended ultra
faint objects, eUFO).

Systems in the same magnitude range, but with smaller sizes ($\reff<20\,$pc),
such as Koposov 1 \& 2 \citep{2007ApJ...669..337K}, Balbinot 1
\citep{2013ApJ...767..101B}, Kim~1 \& 2 \citep{2015ApJ...799...73K,
2015ApJ...803...63K}, are most likely ordinary star clusters, although no
spectroscopic follow-up has been done for any of these objects. 
  
Kinematic data excludes the possibility of large amounts of dark matter in some UFOs (e.g. Segue 3, \citealt{2011AJ....142...88F}), but for others,  the stellar velocities alone do not allow a conclusive classification (e.g. Segue 1, \citealt{2007ApJ...654..897B,2011ApJ...733...46S}). In several cases a spread in $\feh$ is indicative of an extended star formation history and therefore argues for a galaxy classification \citep{2012AJ....144...76W}. 
  However, a prolonged star formation history within a dark matter
halo does not guarantee that the system contains dark matter at the
present day. Tidal stripping and mass segregation could remove the dark matter
halo and leave a dark matter free remnant stellar population orbiting the Milky
Way
\citep{1996ApJ...461L..13M,2005ApJ...619..243M,2005ApJ...619..258M,2008MNRAS.391..942B}.

In the $\Lambda$CDM cosmology \citep{1985ApJ...292..371D,1987ApJ...313..505W,1994ApJ...437L...9C,1996ApJ...462..563N,2006Natur.440.1137S,
2014JPhG...41f3101R}, the smallest galaxies
are believed to have the highest dark matter density and this makes them
promising targets for observing dark matter annihilation signals in
$\gamma$-rays (e.g. \citealt{PhysRevD.89.042001}). Indeed, the Fermi $\gamma$-ray
satellite is observing several UFOs  \citep{2015PhRvL.115h1101G,
2015ApJ...809L...4D}, such as the ones that were recently
discovered in the Dark Energy Survey  data \citep{2015ApJ...807...50B,2015ApJ...805..130K}. 
There is an advantage of looking at the UFOs as opposed to
the Galactic centre, because they contain fewer known $\gamma$-ray sources, such
as radio pulsars and low-mass $X$-ray binaries.  

Uniquely establishing whether a UFO contains dark matter is challenging,
because only a handful of bright stars are available for spectroscopy and
membership determination. 
In addition, it has been proposed that
unbound stars escaping from a dark matter free system could enhance the
velocity dispersion and mimic the effect of a dark matter halo
\citep{1997NewA....2..139K}. For the UFOs, apart from the kinematic challenge, it is also
difficult to determine $\MV$ and $\reff$, which affects the virial mass estimate
because it is proportional to $\reff$. It is, therefore, not inconceivable that
a dark matter free dissolving star cluster appears to have a massive dark
matter halo; this was recently proposed for Segue 1 by \citet{2016arXiv160608778D}. In this paper, we do not focus our study on a particular object, but we aim to shed light on how many star clusters are expected to contribute to the UFO population.

This paper is organised as follows. 
In Section~\ref{sec:Analysis}, we estimate how many faint star clusters (dark matter free objects with $\MV\gtrsim-3.5$ and in the MW-halo) we can expect based on an extrapolation from nearby and bright GCs. 
In Section~\ref{sec:UFO_sim}, we describe the \nbody
simulations to model star clusters. In Section~\ref{sec:results}, we discuss the
results we obtained considering observational biases, and a summary of our 
results is presented in Section~\ref{sec:conclusion}.

\section{The expected number and radius of faint star cluster} 
\label{sec:Analysis}

\subsection{Number of faint star clusters}\label{sec:N_UFO}
In this section we estimate the number of star clusters that are expected to
contribute to the luminosity range of the UFOs by extrapolating
from the known GC population.

We use analytic functional forms for the initial distributions of star cluster
masses and Galactocentric radii, which we then evolve by a simple mass loss
prescription to include the effect of dynamical evolution (two-body relaxation)
in the Milky Way potential.

We assume a Schechter function \citep{1976ApJ...203..297S} for the clusters 
initial mass function (hereafter, CIMF; \citealt{2007ApJS..171..101J} for old
clusters; and \citealt{2006A&A...450..129G} and \citealt{2009A&A...494..539L}
for young cluster):

\begin{equation}
\frac{\dr N}{\dr \Mi}=A\Mi^{-\alpha} \exp\left(-\frac{\Mi}{M_*}\right).
\end{equation}
where $\Mi$ is the initial mass of star clusters, $M_*$ is the mass where the
exponential drop occurs, $A$ is a constant that sets the total mass in clusters 
and $\alpha$ is the power-law index
at low masses $\Mi\lesssim M_*$.

Because the Milky Way GCs are old and have lost mass as the result of dynamical
evolution, we are interested in the {\it evolved} mass function
\citep{2007ApJS..171..101J}, which can be expressed in the CIMF by using
conservation of number \citep{2001ApJ...561..751F}

\begin{equation}
f(M, \RG) = \frac{\dr N}{\dr M}=\frac{\dr N}{\dr \Mi} \left\vert {\frac{\partial \Mi}{\partial M}} \right\vert,
\label{eq:fmrg}
\end{equation}
where $\RG$ is the Galactocentric radius which enters because the mass evolution
depends on the orbit.  To proceed, we need an expression for
$\partial\Mi/\partial M$ that encapsulates the physics of mass loss of GCs. We
assume that the dominant mass-loss process is evaporation, which is the result
of two-body relaxation in the Galactic tidal field. \citet{2001MNRAS.325.1323B}
showed that for this process, the dissolution time-scale of GCs, $\tdis$, scales
with their two-body relaxation timescale, $\trh$, as $\tdis \propto \trh^{x}$,
with $x\simeq3/4$. The mass-loss rate, $\dot{M}$, can then be written as
$\dot{M} = \dot{M}_5(\RG)(M/10^5\,\msun)^{1-x}$, where $\dot{M}_5(\RG) \simeq
20\,\msun\,\Myr^{-1}\,(\kpc/\RG)$  is the $\RG$-dependent mass-loss rate found
in models of a cluster with a mass of $10^5\,\msun$ on a circular orbit in an
isothermal Galactic potential \citep*{2011MNRAS.413.2509G}.  From integrating
$\dot{M}$ we can find an expression for $M(\Mi,\dot{M}_5, \age)$
\citep{2005A&A...441..117L} from which we derive

\begin{align}
\Mi &= \left(M^x + \Delta_x\right)^{1/x}\label{eq:mim}\\
\frac{\partial \Mi}{\partial M} &= M^{x-1} \left(M^x + \Delta_x\right)^{(1-x)/x},\label{eq:partial}
\end{align}
with $\Delta_x = x\,(1-\epsilon)^{-1}\,(10^5\,\msun)^{1-x}\,\dot{M}_5\,(\age/\Myr)$, where $\epsilon$ is the eccentricity of the orbit. The  $(1-\epsilon)^{-1}$ term encapsulates the fact that clusters on eccentric orbits lose mass faster \citep{2003MNRAS.340..227B,2016MNRAS.455..596C}. We adopt $\epsilon =0.5$, which corresponds to the typical eccentricity of isotropic orbit distribution in a singular isothermal sphere \citep{1999ApJ...515...50V}.
Combining equations~(\ref{eq:fmrg}), (\ref{eq:mim}) and (\ref{eq:partial}) we find an expression for the evolved clusters mass function \citep{2009MNRAS.394.2113G}
\begin{equation}
f(M, \RG)= A \ \frac{M^{x-1}}{\left( M^x + \Delta_x \right)^{\frac{\alpha+x-1}{x}}}
\ \exp\left(-\frac{\left( M^x + \Delta_x \right)^{1/x}}{M_*}\right).
\label{eq:fmrg2}
\end{equation}

Because we are interested in finding how many faint star cluster (dark matter free objects with $\MV\gtrsim-3.5$ and $\RG\geq20\,\kpc$, hereafter FSC) we expect in the outer halo of the Milky Way, we need to adopt a Galactocentric radius distribution. We decide to use a simple power-law for the initial distribution 
\begin{equation}
g(\RG) = \left.\frac{\dr N}{\dr \RG}\right\vert_i = \RG^{2-\beta},
\end{equation}
where $-\beta$ is the index of the number density distribution $n(\RG)$, because $g(\RG) = 4\pi \RG^2n(\RG)$.

The bivariate distribution that we can compare to the data is thus
\begin{equation}
h(M, \RG) = \frac{\dr^2 N}{\dr M \dr\RG} = f(M,\RG) \ g(\RG),
\label{eq:h}
\end{equation}
where $A$ in the function $f(M,\RG)$ (equation~\ref{eq:fmrg2}) is a constant that sets the number of clusters after integrating $h(M,\RG)$ over $M$ and $\RG$. 
This function can now be used to do a maximum likelihood fit to find the set of free parameters for which the distribution ($h(M,\RG)$ in our case) becomes most probable:

\begin{equation}
\ln \mathcal{L} = \sum_i \ln \ell_i(p_1,p_2,...,p_j)
\end{equation}

where $\ell_i(p_1,p_2,...,p_j)$ is the probability of finding the datum $i$ given the set of parameters $p_1,p_2,...,p_j$.
In our case:

\begin{equation}
\ln \mathcal{L} = \sum_{i=1}^{\nsel} \ln \left[ h_i \left( \alpha, \beta	,x, M_* \right)  \right],
\end{equation}
where $h_i = h(M_i, {\RG}_i)$ and $\nsel$ is the number of clusters in the sample.

\begin{figure}
\center
\includegraphics[width=0.48\textwidth]{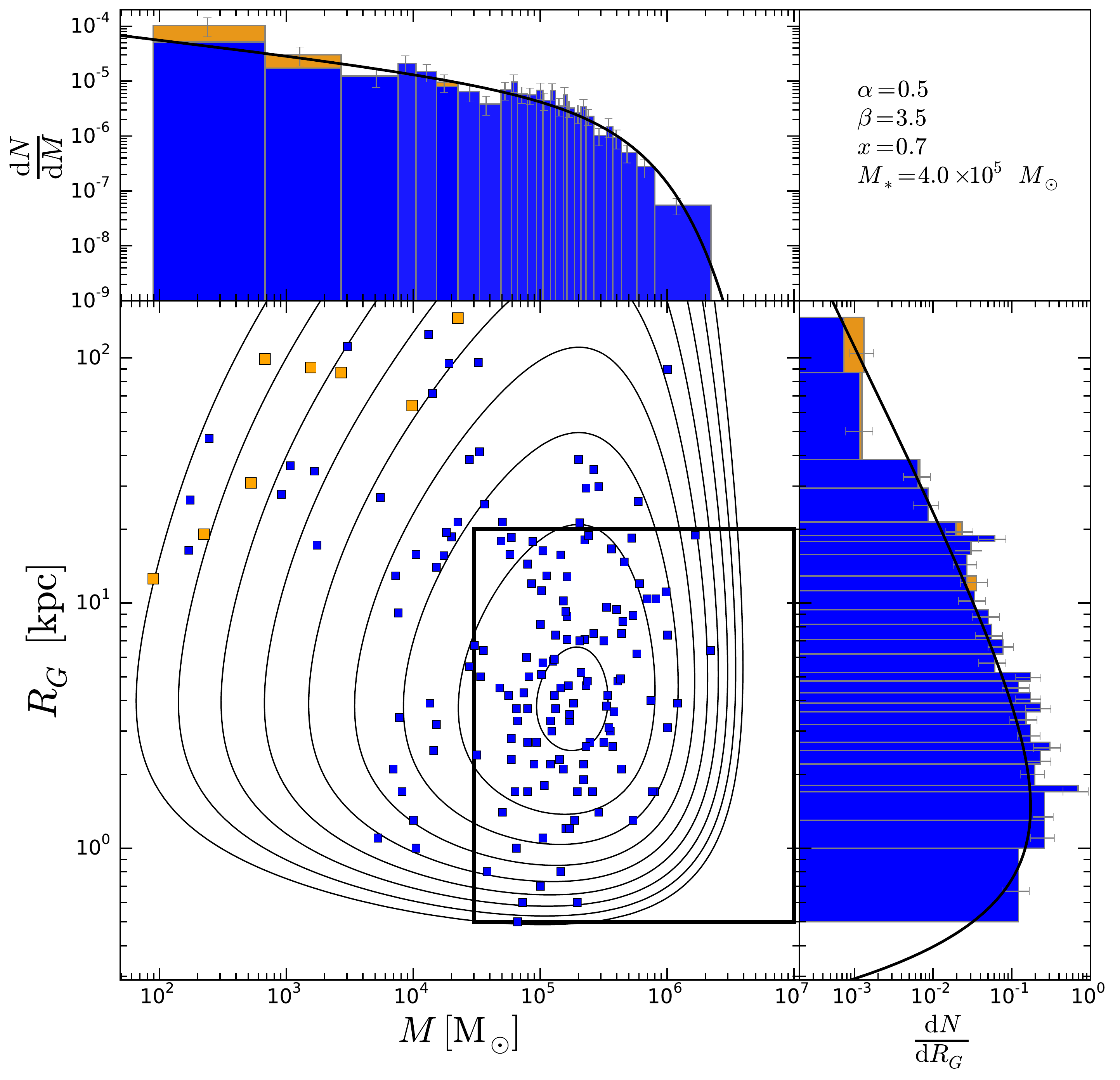}
\caption{In the bottom left plot we show the Milky Way GCs (blue squares) and the GCs candidates discovered in the last three years (orange squares): Laevens 1/Crater, Laevens 3, Kim 1, Kim 2, Eri III, Balbinot 1, DES 1, Kim 3. The area in the black box is where we compute our fit.
In the upper left plot, we have the normalized mass function versus the mass of the GCs, while in the bottom right there is the normalized distribution function versus the Galacticentric distance of the GCs. In blue, the histogram for all the GCs, while the best fit (black line) with $\alpha$, $\beta$, $x$ and $M_*$ as parameters calculated with {\sc EMCEE}, was found selecting the GCs in this region: $0.5<\RG /\kpc<20$ and $3 \times 10^4<M/\msun< 10^7$. The results for the parameters are shown in the upper right plot. In the histograms, the error bars (in grey) are estimated using a Poisson error.
}\label{fig:dNdMdRG}
\end{figure}

We use the \citet{2010arXiv1012.3224H} catalogue of Milky Way globular cluster properties to get $M$ and $\RG$ for each cluster and use $\ml=2$ to convert luminosities to masses. 
 We then use a Monte Carlo Markov Chain (MCMC) method (the affine-invariant ensemble sampler as implemented in the {\sc emcee} code, \citealt{2013PASP..125..306F})  to find the parameters: $\alpha$, $\beta$, $x$ and $M_*$ that give the highest likelihood. We decide to fit equation~(\ref{eq:h}) to the GCs in the $M$ range $ 3\times 10^4<M/\msun <10^7$ and $\RG$ range $0.5<\RG/\kpc<20$, because this is where we believe the catalogue is complete. The number of selected GCs in that range is $\nsel = 115$.
 In Tab.~\ref{tab:fit_param}, we show the results of our best fit parameters. In Fig.~\ref{fig:dNdMdRG} we show the resulting best-fit distribution.

We then use the best fit distribution to estimate the number of low-mass GCs at large $\RG$, where the Harris catalogue is incomplete. With the known parameters of the $h(M, \RG)$ distribution, it is possible to estimate the number of faint star clusters ($N_{\rm FSC}$) by integrating the distribution over the range where the known UFOs are found $\left( 20\leq{\RG}/\kpc\leq150; \ 10^2\leq M/\msun\leq 4.3 \times 10^3 \right)$.
The lower and upper limit of the mass range correspond to $M_V \simeq 0$ and $M_V \simeq -3.5$, respectively, with our adopted $\ml=2$.

\begin{figure}
\center
\includegraphics[width=0.48\textwidth]{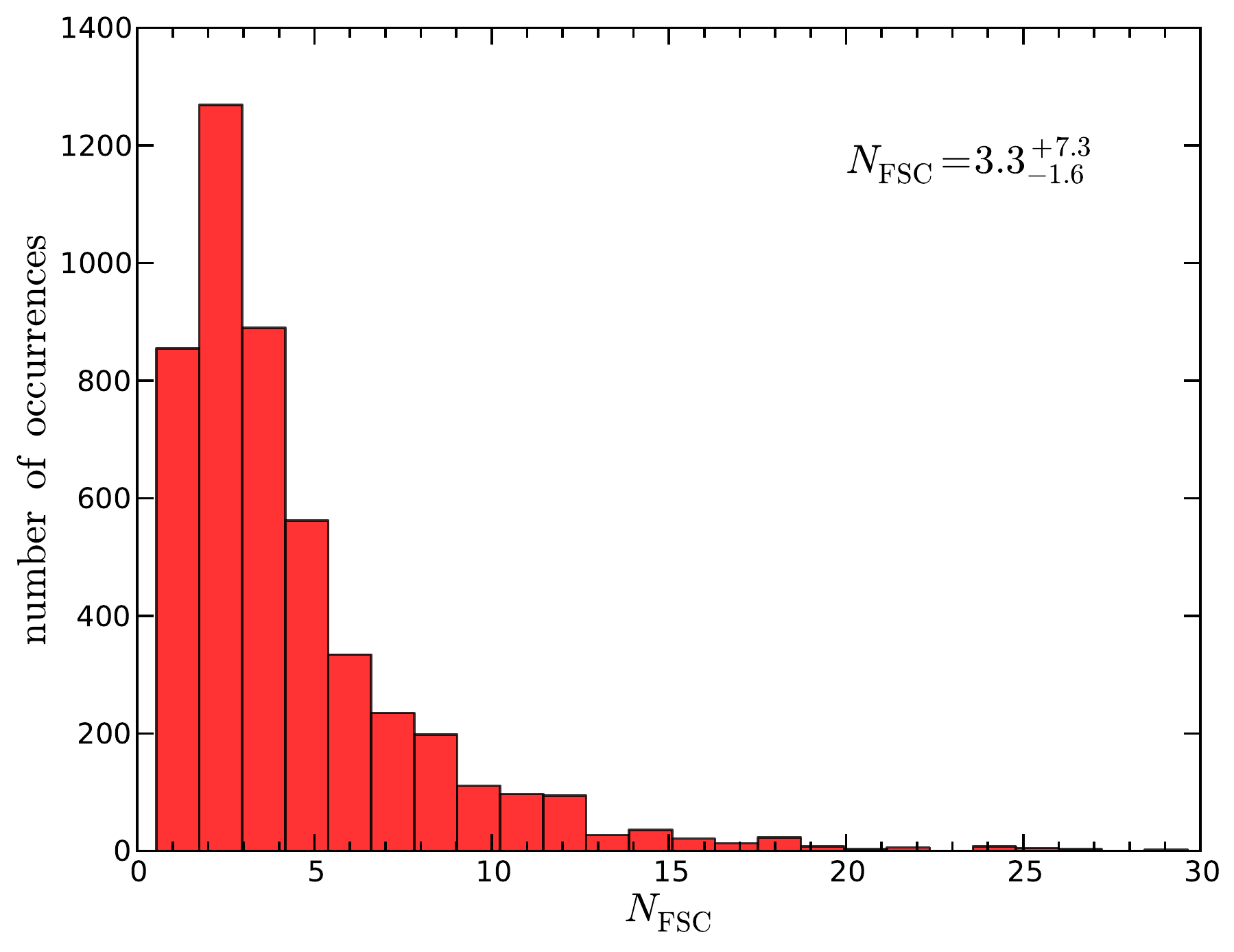}
\caption{Posterior of the number of faint star clusters ($N_{\rm FSC}$) marginalised over $\alpha$, $\beta$, $x$ and $M_*$. The inferred $N_{\rm FSC} = 3.3^{+7.3} _{-1.6}$.  }\label{fig:posterious}
\end{figure}

Therefore, the number of faint star clusters is
\begin{align} 
N_{\rm FSC}
&=
\nsel  \int_{10^2\,\msun}^{4.3 \times 10^3\,\msun} \! \int_{20\,\kpc}^{150\,\kpc} h(M,\RG) \ 
\dr \RG \dr M \nonumber \\
&=3.3^{+7.3}_{-1.6} \label{eq:N_FSC}
\end{align}
where the constant $A$ in $h(M, \RG)$ is such that an integration over the range used for the fit results in 1. 
The quoted value is the median of posterior distribution of $N_{\rm FSC}$ shown in Fig.~\ref{fig:posterious}, and the uncertainties correspond to the region containing 68.3\% of the points around the median. 

In Fig.~\ref{fig:dNdMdRG} we show that the extrapolation from the fit to the bright GCs agrees with the number of observed cluster with and without the last observed GC candidates (in orange); however based on this we cannot conclude that a fraction of UFOs need to be galaxies.

\begin{table}
\center
\begin{minipage}{60mm}
\caption{Best fit parameters
}
\label{tab:fit_param}
\resizebox{\linewidth}{!}{

\begin{tabular}{@{}lcc}
\hline
\\[-2ex] 
Parameter & Value & Unit\\
\\[-2ex] 
\hline
\\[-2ex] 
$\alpha$ & $0.452\pm0.236$ & \\
$\beta$ & $3.523\pm0.128$ &\\
$x$ & $0.724\pm0.090$ &\\
$M_*$ & $4.041\pm0.964$ & $10^5\,\msun$\\
\\[-2ex]
\hline
\end{tabular}}
\medskip

\end{minipage}
\end{table}

\subsection{Size estimate of faint stellar systems}\label{sec:estimation size} 
The UFOs have sizes up to approximately $100\,\pc$ (Fig.~\ref{fig:lumVSradius}), but the uncertainties can sometimes be extremely large. \citet*{2012ApJ...745..127M} show that it can be challenging to estimate the structural parameters of the ultra-faint DGs within 10\% of their true value.
For star clusters, it is not known whether it is possible that they appear that large. 
Here we  estimate the maximum radius that a star cluster can have,  which corresponds to the situation in which the cluster fills the Roche volume. In that case the half-mass radius ($\rh$) depends on the strength of the tidal field along the orbit. 
As described by \citet{H61}, a star cluster evolving in a tidal field,
 evolves at a constant mean density once it fills the Roche-volume, which means that the ratio between the $\rh$ and the Jacobi radius ($\rJ$) is constant: $\rh/\rJ \simeq0.15$.
This fraction is somewhat $N$-dependent \citep{2012MNRAS.422.3415A} and can be as large as $\rh/\rJ \simeq 0.4$ \citep{1997MNRAS.286..709G} for very small $N$, i.e. our region of interest.
The Jacobi radius is defined in \citet{1962AJ.....67..471K} as:

\begin{equation}\label{eq:King formula}
\rJ =\left(\frac{G M}{{\Omega}^2-\frac{\partial^2\phi}{\partial \RG^2}}\right)^{\nicefrac{1}{3}}, \qquad
\end{equation}
where $\Omega$ is the angular velocity of the cluster around the Galaxy centre, 
 $\phi$ is the potential of the Galaxy and  
 $G$ is the gravitational constant. 
Therefore, using equation~(\ref{eq:King formula}) and assuming a singular isothermal halo we obtain 
\begin{equation}\label{eq:isothermal}
\rJ =\left(\frac{G M}{{2\Omega}^2}\right)^{\nicefrac{1}{3}}, \qquad  
\end{equation}
which is only valid for circular orbit, thus $\Omega=V_{\rm C}/\RG$.

Therefore, using $\rh/\rJ = 0.2$ for a cluster with $M = 500\,\msun$ at $\RG = 50\,\kpc$, we find $\rh \simeq 7.6\,\pc$ (with \mbox{$\reff\simeq5.7\,\pc$}, if we assume that mass follows light).

From this we see that the radii of the compact UFOs are consistent with being
tidally limited star clusters. However, star clusters in the end of their life
have lost most of their low-mass stars, and will therefore have a smaller $M$,
resulting in smaller a $\rJ$ and hence a smaller $\rh$. Also, if the cluster is mass
segregated, $\reff$ can be smaller than $\rh$ in projection
\citep{2007MNRAS.379...93H}. However, observational biases, such as the presence of
unbound stars and dark remnants \citep{2016MNRAS.462.2333P} could perhaps inflate $\reff$ with respect to $\rh$. In the next
section we consider the evolution of $\reff$ in numerical models, taking all
observational biases into account.

\section{Numerical N-body simulations of FSC{\scriptsize s} }\label{sec:UFO_sim}

\subsection{Description of the $N$-body simulations}\label{subsec:galactic_potential}
In this section we describe the details of the simulations. 
In order to simulate the evolution of star clusters in a tidal field, we used {\sc nbody6tt} \citep{2011MNRAS.418..759R,2015MNRAS.448.3416R}, which is an adaptation of the widely used direct $N$-body code {\sc nbody6} developed by \citet{2012MNRAS.424..545N}. We use {\sc nbody6tt} because we want to consider a Galactic potential that is currently not available in {\sc nbody6}.
With {\sc nbody6tt} it is straightforward to include the tidal field due to an external galactic potential  that is a function of position and time. The tidal force is added to the equation of motion of a star in a non-rotating frame by adding the difference in galactic acceleration on the star and the guide centre.  The guide centre is a pseudo-particle (initially at the centre of mass of the cluster), and its motion is integrated separately \citep{2003gnbs.book.....A}.  

We adapt three different Galactic potentials: a static `NFW-potential' \citep*{1996ApJ...462..563N}, a `growing NFW-potential' (hereafter gNFW, \citealt{2014A&A...563A.110B}), and a three component potential \citep{1990ApJ...348..485P}.

34 simulations were performed using the static NFW potential:
\begin{equation}\label{eq:NFW}
\phi_{\rm NFW} = -\frac{GM_0}{\RG}\ln\left(1+\frac{\RG}{R_0}\right)
\end{equation}
where the scale mass $M_0$ is chosen to have a maximum circular velocity  $\VC = 210\,\kms$ at $\RC = 30\,\kpc$, and the scale radius $R_0$ is $13.9\,\kpc$.

To test the role of the Galactic potential, 8 simulations were performed 
using the analytical gNFW model of \citet{2014A&A...563A.110B}:
\begin{equation}
\phi_{\rm gNFW} = -\frac{G \Ms(z)}{\RG}\ln\left(1+\frac{\RG}{\Rs(z)}\right)
\end{equation}
where the scale mass ($\Ms$) and the scale radius ($\Rs$) evolve with the redshift $z$ as:
\begin{align*}
	\Ms(z)=M_0 \exp(-0.2z)\\
	\Rs(z)=R_0 \exp(-0.1z)
\end{align*}
with $M_0=\Ms(z=0)$ and $R_0=\Rs(z=0)$ (same values of eq.~\ref{eq:NFW}).

An additional 4 simulations were performed using a three component potential (bulge, disc and halo). We 
used the analytical model from
\citet{1990ApJ...348..485P} (hereafter, P90). 

Bulge:\\
\begin{equation}
\phi_{\rm b} = -\frac{G\Mb}{\sqrt{R^2 + \left(\ab + \sqrt{z^2 + {\bb}^2}\right)^2}}
\end{equation}\\
where: $R$ is the Galactocentric distance in the $x$-$y$ plane; $z$ is the Galactocentric distance in the $z$-component; $\Mb= 6.15\times 10^9 \,\msun$; $\ab= 0.0\,\kpc$; and $\bb =0.277\,\kpc$.\\
\\
Disc:\\
\begin{equation}
\phi_{\rm d} = -\frac{G\Md}{\sqrt{R^2 + \left(\ad + \sqrt{z^2 + {\bd}^2}\right)^2}}
\end{equation}\\
where: $\Md=4.47\times 10^{10} \,\msun$, $\ad= 3.7\,\kpc$, and $\bd =0.20\,\kpc$.\\
\\
Halo:\\
\begin{equation}
\phi_{\rm h} = \frac{G\Mh}{d} \left[\frac{1}{2}\ln\left(1+\frac{\RG^2}{d^2}\right)+\frac{d}{\RG}\arctan\frac{\RG}{d}\right]
\end{equation}\\
where: $\Mh=3.38\times 10^{10} \,\msun$ and $d= 6.0\,\kpc$.\\

In this paper, we choose different values of the masses for different components with 
respect to the ones from \citet{1990ApJ...348..485P}. This difference is due to
a rescaling factor, such that the NFW and P90 galaxy models have the same virial mass. We used the ratio
 between the virial mass of the NFW model ($M_{\rm vir,NFW}=1.26\times10^{12}\,\Msun$) and the virial mass of the original P90 model ($M_{\rm vir,P90}=2.29\times10^{12}\,\Msun$) to rescale the virial mass of the three components in P90. Once the new virial mass of the single components are known it is possible to derive the new $\Mb$, $\Md$ and $\Mh$. The virial mass is the mass of the galaxy within the virial radius, when the mean density of the galaxy is equal to $200\rho_{\rm c}$, where $\rho_{\rm c} = 3H^2_0 / 8 \pi G$ is the critical density and $H_0=68.0\,\kms \, \rm Mpc^{-1}$ is the Hubble constant.
 
Depending on the orbit and the Galactic potential, stars escape from the cluster as result of two-body relaxation. We therefore need to find the initial $N$ that results in near dissolution (i.e. a few bound stars left) at an age of $12$ Gyr. We used the fast cluster evolution code {\sc emacss} \citep{2014MNRAS.442.1265A} to iteratively find the initial $N$ that satisfies these constraints. 
We consider both circular and elliptical orbits for the clusters, with eccentricities of $\epsilon=0$, $\epsilon = 0.25$, $0.5$ and $0.75$  and with apogalactic distances of $50\,\kpc$, $100\,\kpc$ and $150\,\kpc$. In the P90-potential, the apocentre of the clusters were chosen such that the orbits are not planar. 
Escapers were not removed from the simulations to allow stars to move from the tidal tails back into the region of the cluster because of compression at apocentre. 
For the initial conditions of all our clusters we used a Plummer model \citep{1911MNRAS..71..460P} with two different initial densities:  
clusters that are initially Roche-filling  (the stars occupy the total tidal volume), with $\rh/\rJ=0.1$;
 and clusters that are initially Roche-underfilling  (the stars occupy the central region of the tidal volume), where the density within $\rh$ is 
$\rhoh =10^4 \, \msun \, \pc^{-3}$.
The stars in the cluster initially follow  a Kroupa IMF \citep{2001MNRAS.322..231K} between $0.1\,\msun$ and $100\,\msun$, and a  metallicity of $Z=0.0008$ (corresponding to $\feh\simeq-1.5$). 
Moreover, for 7 simulations, we consider the possibility that BHs do not receive a natal kick 
when they form; as a consequence, we retain 100\% of stellar mass black hole initially. While in 
the other simulations the BHs receive a natal kick velocity which is the same kick 
velocity given to the neutron stars.

Furthermore, in some models primordial binaries were included, where the binaries components have the same mass. We used the description by
\citet{1995MNRAS.277.1507K}, where the eccentricities are in thermal distribution with eigenevolution. The distribution of the semi-major axis is either derived from the period distribution, which is initially
an uniform $\log$-period distribution or an uniform distribution for the log of the semi-major axis. 
The properties of the simulations are presented in Table~\ref{tab:UFO_param}. 

\begin{singlespace}
\begin{table}
\scriptsize
\center
\begin{minipage}{80mm}
\caption{\nbody simulation properties
}
\label{tab:UFO_param}
\resizebox{\linewidth}{!}{
\begin{tabular}{@{}lcccc}
\hline
Model & $R_{\rm apo}$ & $\epsilon$ & $N$ & $N_{12 \,\Gyr}$ \\
  \ & $\left[\kpc\right]$ & \ &\\
\hline
\hline
NFW potential\\
\hline
50e00H & 50 & 0.00 & 4096 & 240  \\
50e25H & 50 & 0.25 & 5000 & 162  \\
50e50H & 50 & 0.50 & 6000 & 184  \\
50e75H & 50 & 0.75 & 10000 & 147 \\
\hline
50e00L & 50 & 0.00 & 2048 & 67 \\
50e25L & 50 & 0.25 & 3000 & 44 \\
50e50L & 50 & 0.50 & 8192 & 180  \\
50e75L & 50 & 0.75 & 20000 & 91  \\
\hline
100e00H & 100 & 0.00 & 2048 & 212 \\
100e25H & 100 & 0.25 & 2048 & 87  \\
100e50H & 100 & 0.50 & 3000 & 217  \\
100e75H* & 100 & 0.75 & 3000 & 13 \\
\hline
100e00L & 100 & 0.00 & 1024 & 125  \\
100e25L & 100 & 0.25 & 1024 & 71 \\
100e50L & 100 & 0.50 & 2048 & 159  \\
100e75L* & 100 & 0.75 & 8192 & 29 \\
\hline
150e00H & 150 & 0.00 & 1500 & 172 \\
150e25H & 150 & 0.25 & 2048 & 246  \\
150e50H & 150 & 0.50 & 1500 & 44 \\
150e75H & 150 & 0.75 & 2048 & 53 \\
\hline
150e00L & 150 & 0.00 & 512 & 113 \\
150e25L & 150 & 0.25 & 1024 & 211  \\
150e50L & 150 & 0.50 & 1500 & 183  \\
150e75L & 150 & 0.75 & 2048 & 82 \\
\hline
50e50M-B1 & 50 & 0.50 & 7200 & 182 \\
50e50M-B2 & 50 & 0.50 & 7200 & 227 \\
50e50M    & 50 & 0.50 & 6000 & 193 \\
\hline
50e50H-BH  & 50 & 0.50 & 6000 & 164  \\
50e50L-BH* & 50 & 0.50 & 30000 & 250  \\
50e50L-B2-BH* & 50 & 0.50 & 30000 & 0 \\
50e75H-BH & 50 & 0.75 & 10000 & 212 \\
50e75L-BH$^{\dagger}$ & 50 & 0.75 & 32768 & 32  \\
150e25H-BH & 150 & 0.25 & 2048 & 139  \\
150e25L-BH & 150 & 0.25 & 1200 & 176  \\
\hline
\hline
gNFW potential \\
\hline
50e50H-g & 50 & 0.50 & 6000 & 135  \\
50e50L-g & 50 & 0.50 & 5000 & 71  \\
50e75H-g  & 50 & 0.75 & 8192 & 89 \\
50e75L-g* & 50 & 0.75 & 10000 & 90  \\ 
150e25H-g & 150 & 0.25 & 1500 & 120  \\
150e25L-g & 150 & 0.25 & 1024 & 253  \\
150e75H-g & 150 & 0.75 & 2048 & 66 \\
150e75L-g & 150 & 0.75 & 2048 & 149 \\
\hline
\hline
P90 potential \\
\hline
50e50H-P90 & 50 & 0.50 & 5000 & 110\\
50e50L-P90 & 50 & 0.50 & 4096 & 177  \\
50e75H-P90  & 50 & 0.75 & 10000 & 166 \\
50e75L-P90* & 50 & 0.75 & 17000 & 159  \\
\hline
\end{tabular}
}

\medskip
\footnotesize{Note. --- The capital letter in the model label indicates if the model was, as initial condition, underfilling (high density, H) or Roche-filling (low density, L). In column 4 we show the initial number of stars; column 5 are the number of bound stars at $12\,\Gyr$. The models with the letter M are simulations with a different initial density ($\rho_{\rm h}=10^3\,\msun/\pc^3$), with B1 and B2 contains $\sim20\%$ of primordial binaries, but different semi-major axis distributions; and with BH retain 100\% of BHs initially. In gNFW the value of $\epsilon$ is the eccentricity at $\sim12\,\Gyr$. The star (*) and the $\dagger$ denote models for which $\rh/\rJ = 0.09$ and $\rh/\rJ = 0.06$ respectively, i.e. slightly denser to avoid a high escape rate on a dynamical time.}
\end{minipage}
\end{table}
 \end{singlespace}
 
\subsection{Model for the background stars}\label{sec:bg} 

Typically, observers use simple colour-magnitude cuts to select cluster stars
with respect to a fore/background. This method, however, does not completely
eliminate the contamination from Milky Way field stars. In order to
account for this issue we adopt a synthetic Milky Way stellar population. We used
the code \trilegal\ 1.6 \citep{2012rgps.book..165G}, which models the Milky Way
stellar population for a given region in the sky, we created a map of stars at
two positions ${ (\ell, b)=(158.6°, 56.8°)}$; and ${ (\ell, b)=(260.98°,
70.75°)}$. The simulated backgrounds are at the positions of the known UFOs, Koposov 1 (Ko1, \citealt{2007ApJ...669..337K}); and Willman 1 (Wil1,
\citealt{2005AJ....129.2692W,2006astro.ph..3486W,2011AJ....142..128W}).  Ko1 has
a small half-light radius $\sim3$ pc, while Wil1 has a large half-light radius
$\sim25$ pc, which are extremes in size for this class of objects. Our goal is
to see whether a cluster with a different background star density can appear
bigger or smaller. 

The \trilegal\ sample was created assuming literature values for the reddening
\citep{1998ApJ...500..525S}.  Assuming $R_V=3.1$ (typical for the Milky Way) and
a calibration at infinity, we obtain an extinction of $A_V(\infty)=0.0418$ for
Wil1 and $A_V(\infty)=0.0757$ for Ko1, which is used by \trilegal\ to simulate
extinctions which are normally distributed. The scatter on the extinction is
also taken from \citet{1998ApJ...500..525S} dust maps. 

In order to introduce some noise in the reddening correction we proceed to
correct the \trilegal\ sample assuming a single average value of extinction for
the full simulated region. 
This adds uncertainty to the reddening, which is likely to be the case in real
observations.  

Furthermore, we assume a photometric error curve $\nu$, with
an exponential form, which represent a typical error in mag for each star.

Here the steps to estimate the background number density:

\begin{enumerate}
	\item Correction for extinction:
	\begin{align*}
	g'=g-A_g \\
	r'=r-A_r
	\end{align*}
	where $g$ and $r$ are  apparent magnitudes  in SDSS filters and $g'$ and $r'$ are the extinction corrected equivalents.
	For Ko1: $A_g=0.091$ and $A_r=0.066$; whereas for Wil1: $A_g=0.013$ and $A_r=0.034$. These values are estimated using \citet{1989ApJ...345..245C} and \citet{1994ApJ...422..158O} extinction curve with $R_V=3.1$.
	\item Using a photometric error curve:
	\begin{equation}
	\nu(m,a,b,c)=a+e^{\frac{m-b}{c}}
	\end{equation}
	where $m$ is the observed magnitude corrected for the extinction and $(a,b,c)$ are parameters which depend on the observations; we compute the magnitudes with simulated errors:
	\begin{align*}
	g''=g'+\chi\ \nu(g',a,b,c) \\
	r''=r'+\chi\ \nu(r',a,b,c)
	\end{align*}
	where $\chi$ is a random number sampled from a Gaussian distribution with mean 0 and variance 1.    	For Ko1 we use $(a,b,c)=(0.005,22,1.2)$; whereas for Wil1, $(a,b,c)=(0.005,25,1.2)$. 
	We choose the value of $b$ to match the limiting magnitude of the observations 
	(\citealt{2007ApJ...669..337K} for Ko1 and \citealt{2006astro.ph..3486W} for Wil1).
\end{enumerate}
We use the above procedure for each star, created with \trilegal\ 1.6,
 in the field of view of 3 degree, centred in the position of Ko1 and Wil1 .

Finally, we applied the following colour-magnitude cuts: 
$16 \leq r'' \leq 22$ and $g''-r'' \leq 1.2$ for Ko1; while $22.6<r''<24.8$ and $0.25<g''-r''<0.65$ 
for Wil1; taking into account only the stars that follow these criteria, we can derive the number of stars per arcsec$^2$.

\subsection{Maximum likelihood method to fit half-light radii}\label{sec:radii}
To estimate the $\reff$ of the simulated clusters, we used a maximum likelihood fit following the procedure outlined in \citet{2008ApJ...684.1075M}. 
Having the position of the stars on the plane of the sky, the maximum likelihood fit can find the set of free parameters for which the observations become most probable.

We choose a likelihood ($\mathcal{L}$) in the following form:

\begin{equation}
\ln \mathcal{L} = \sum \ln \left(\nP+\nBG \right)
\end{equation}

where $\nP$ and $\nBG$ are the probabilities of a star belonging to the cluster
and background, respectively. We choose $\nP$ to be a 2-D elliptical Plummer
profile, given by:

\begin{equation}
\nP= \frac{N_*}{(1-e) \pi a^2} \left(1+\frac{d^2}{a^2}\right)^{-2}
\end{equation}

with

\begin{equation}
d^2=\left[\frac{1}{1-e}\left(x\cos(\theta)-y\sin(\theta)\right)\right]^2+
\left[x\sin(\theta)+y\cos(\theta)\right]^2
\end{equation}

In our likelihood analysis we choose the following parameters: the scale
radius ($a$) which is also the projected half-number radius,  the number of
stars in the cluster ($N_*$), the ellipticity\footnote{The ellipticity is defined as $e=1 - b_0/a_0$ where $b_0$ and $a_0$ are the semi-minor and semi-major axis of the ellipse, respectively.} ($e$) and the position angle
($\theta$); while $x$ and $y$ are the positions of the stars on the $x$-$y$ plane.
We can estimate the number of stars in the background $N_{\rm BG}$, fitting on
the parameter $N_*$, and, knowing the number of stars in our snapshot $N_{\rm
tot}$ ($N_{\rm tot}=N_* + N_{\rm BG}$). Therefore, knowing the area of our
simulated field of view, we can derive $\nBG$, which is considered to be
homogeneous across the simulated field-of-views. 
We use a downhill simplex method \citep{citeulike:3009487} to find the
parameters that maximizes the likelihood. 
In the following Section we discuss the results of our analysis.

\section{Results}\label{sec:results}
In this section we present the results from our analysis, discussing the importance of each observational bias. In this way a comparison between \nbody simulations and observational data  can tell us something  about the underlying properties of the observed objects.

\subsection{Example of the evolution of a low-$N$ cluster}
\begin{figure}
\center
\includegraphics[width=0.48\textwidth]{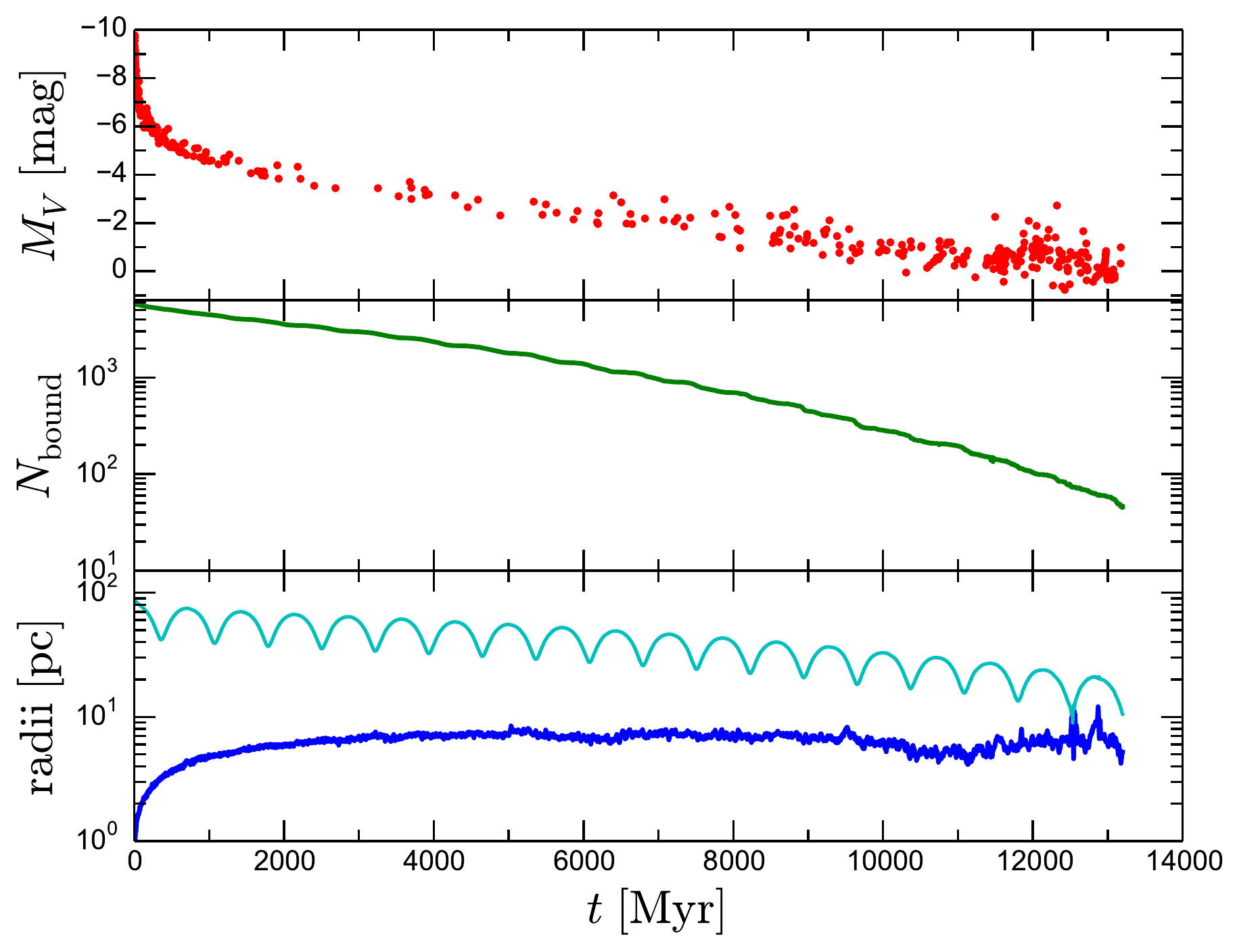}
\caption{
Simulation of a star cluster with initial $N=6000$, apogalaction at $50\,\kpc$ and eccentricity $\epsilon=0.5$. Top: evolution of the absolute magnitude in the $V$-band of all the observable particles with  $r<300\,\pc$.  Middle: evolution of the number of bound stars (green line). Bottom: evolution of half-number radius. The blue line is the half-number radius of the bound stars; the cyan line is the tidal radius: in the local minima the cluster is in pericentre, while in the local maxima the cluster is in apocentre.
} \label{fig:ThreePlots}
\end{figure}

To illustrate the evolution of the underlying cluster properties we first present some of the results without considering observational biases.  
 In Fig.~\ref{fig:ThreePlots} we show the properties of the 50e50H model (see Table~\ref{tab:UFO_param}).
The upper panel shows  the evolution of the absolute $V$-band magnitude ($\MV$, see Appendix~\ref{app:MV} for more details on how $\MV$ has been computed) and from this it can be seen that already at approximately $4\,\Gyr$  the cluster reaches a luminosity of typical UFOs (see Fig.~\ref{fig:lumVSradius}).
From then onwards, until the end of the evolution the total luminosity drops by a factor of $\sim15$, and the cluster remains in the luminosity range of UFOs until complete dissolution. From a comparison to the number of bounds stars ($\nbound$, middle panel), $\nbound$ decreases by a factor of $\sim100$ in this period. The slow decrease in luminosity compared to $\nbound$ is due to mass segregation and the preferential loss of low-mass stars in the late stages of cluster evolution. This means that our previous estimate of the $N_{\rm FSC}$ in the right mass range is a lower limit (eq.~\ref{eq:N_FSC}), because the $N_{\rm FSC}$ in the correct luminosity range is higher. 
In the lower panel of Fig.~\ref{fig:ThreePlots}, we show the evolution of $\rh$ (bottom blue line) and after about $4\,\Gyr$ it levels to a value that is consistent with filling the Roche volume (see  Section~\ref{sec:estimation size}). The top line (cyan) shows the evolution of $\rj$ computed using equation~(\ref{eq:King formula}) which decreases due to the loss of cluster mass because of escaping stars. 

Because some of the above properties, such as $\nbound$, $\rh$ and $\rj$ are not observable, we need to include observational biases in our analyses of the $N$-body results before we can make a meaningful comparison with the observations.  
Therefore, in the next section we analyse our data in a similar way as is done for the observational data, as described in Sections~\ref{sec:bg} and ~\ref{sec:radii}.

\subsection{The effect of the background on the size measurements}\label{sec:bg_effect}
With the procedure explained in Section~\ref{sec:bg}, we include  background stars in our simulations. We derive the number density of the background stars ($\nBG$) for two different observed UFOs: Wil1 and Ko1; where $\nBG({\rm Wil1})<\nBG({\rm Ko1})$. Then, with the cluster in the centre, we add randomly the background stars, uniformly distributed, in an area with a radius of $400\,\pc$, far beyond the tidal radius of a low-mass cluster. Finally, as described in Section~\ref{sec:radii}, we compute the best fit Plummer radii, taking only the `observable stars' into account. We consider `observable stars' all the stars with masses greater than $0.5\,\msun$ and which are not dark remnants.

\begin{figure}
\center
\includegraphics[width=0.48\textwidth]{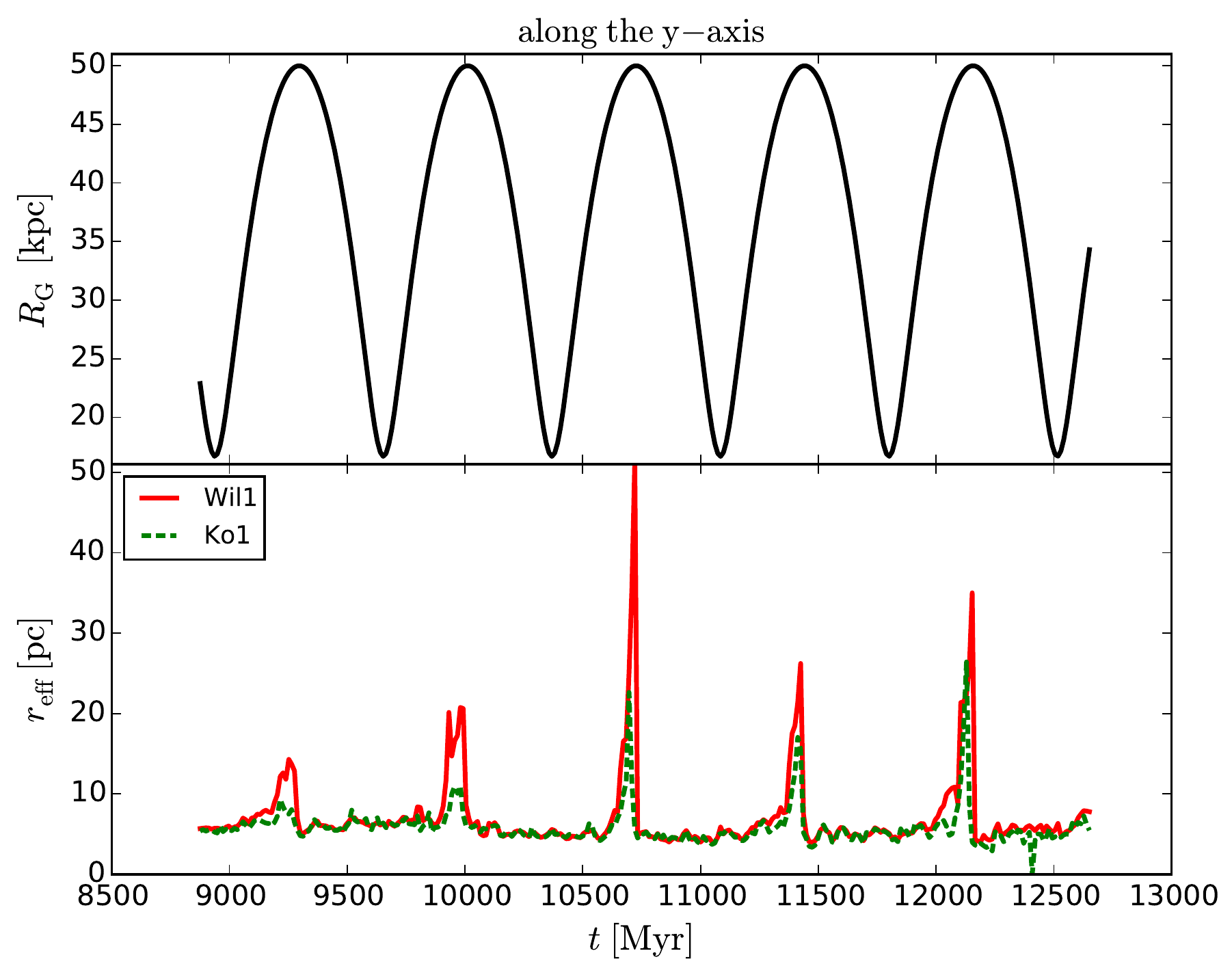}
\caption{
Top: orbit of the cluster.
Bottom: Evolution of the half-number radius of the 50e50H model along the $y$-axis.
Half-number radius for a cluster with a Wil1-like background (red line) and a
Ko1-like background (dashed green line).
}\label{fig:comparex}
\end{figure}

\begin{figure}
\center
\includegraphics[width=0.48\textwidth]{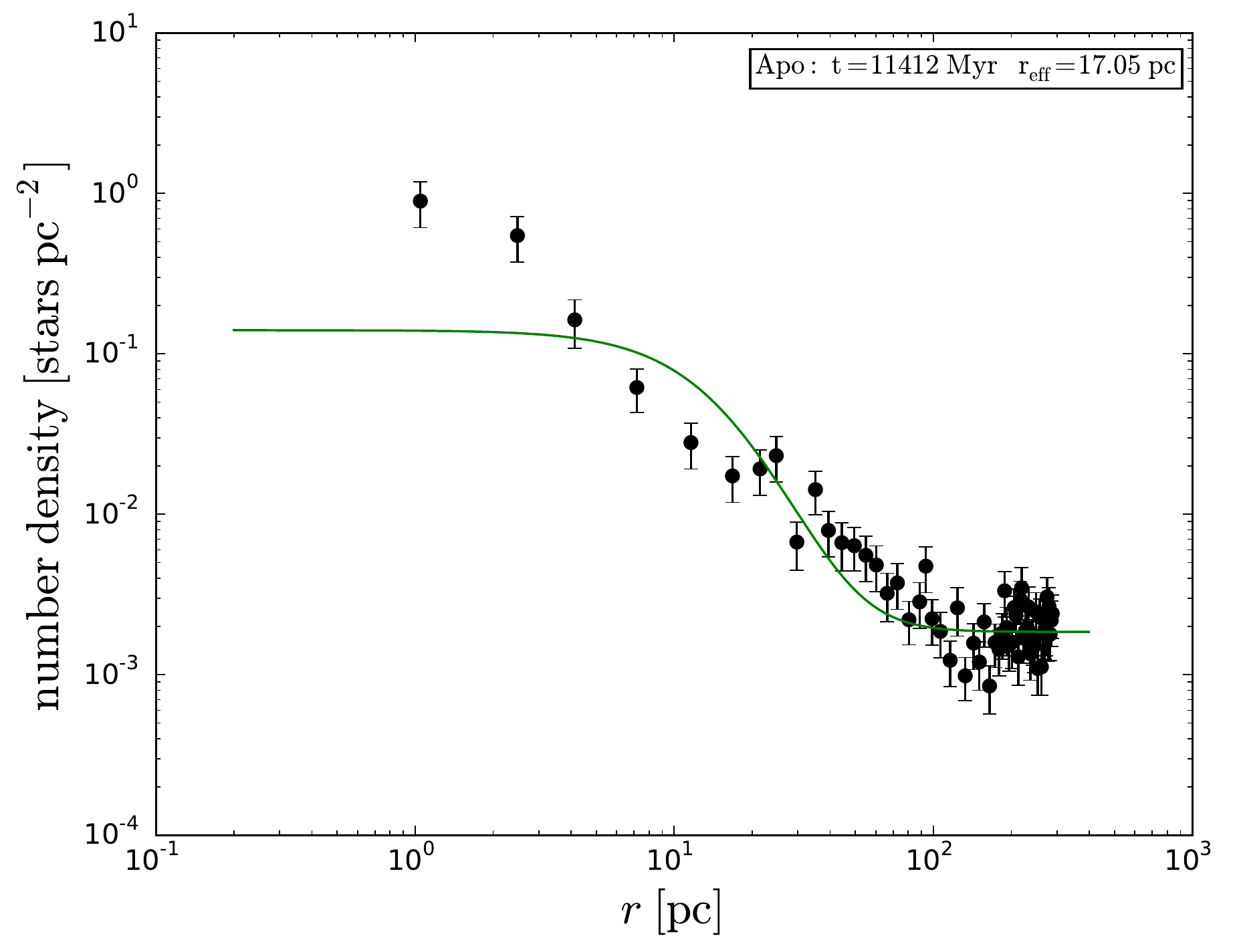}
\caption{Number density profile of the 50e50H model in apocentre along the
    $y$-axis (in this case along the orbit). The green line is the estimated
    Plummer model using the parameters obtained with the maximum likelihood fit.
    	The Plummer model is not ideal to fit this number density profile because,
    at $r\sim25\,\pc$ the bump, caused by the projected positions of the stars in the tails
    which are close to the centre (in a radius of $400\,\pc$), increases the
    estimation of the size.
}\label{fig:sdp_9403}
\end{figure}

\begin{figure}
\center
\includegraphics[width=0.48\textwidth]{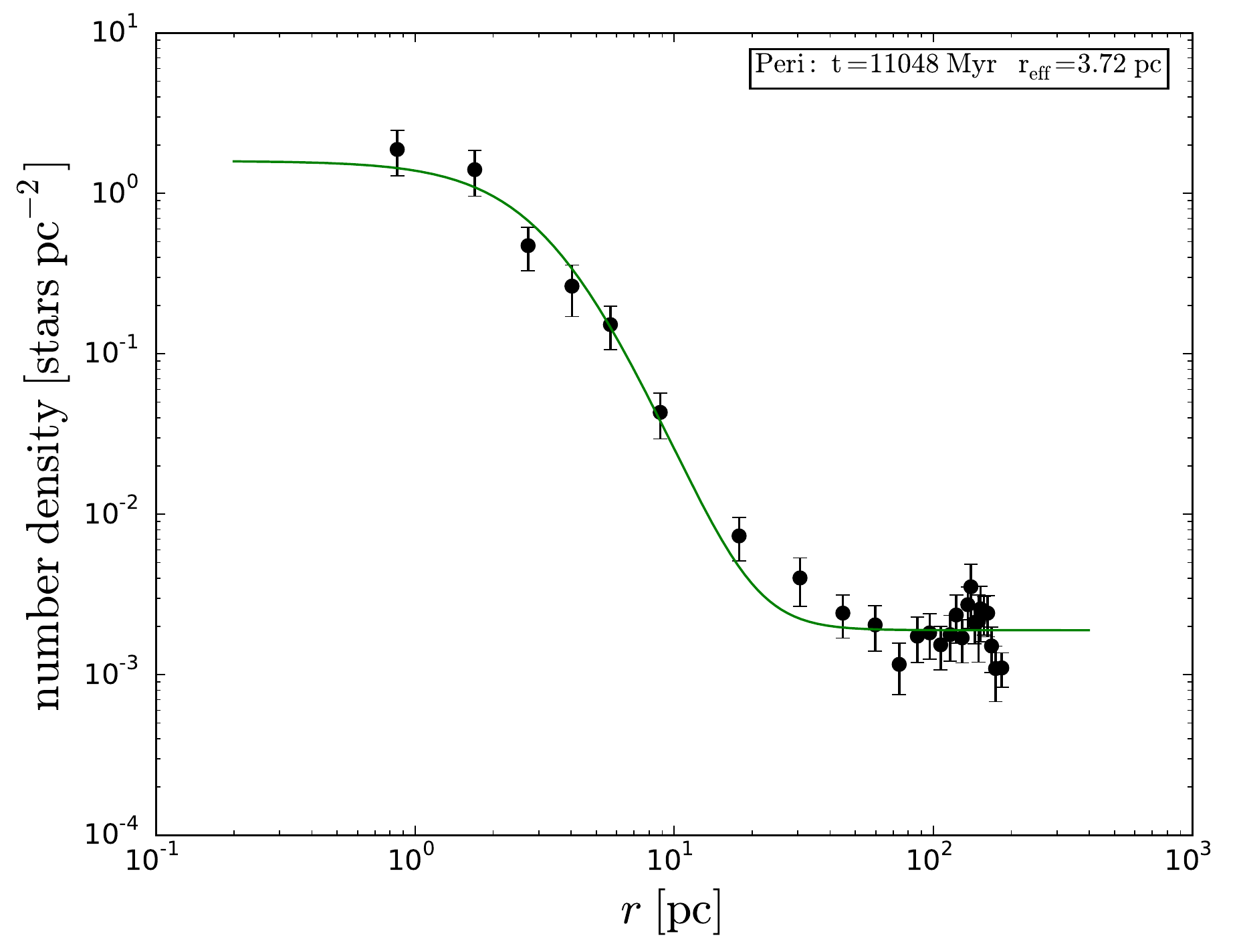}
\caption{Number density profile of the 50e50H model in pericentre along the
    $y$-axis (in this case along the orbit). The green line is the estimated
    Plummer model using the parameters obtained with the maximum likelihood fit.
}\label{fig:sdp_9103}
\end{figure}

In Fig.~\ref{fig:comparex}, we show the evolution of $\reff$ for the model 50e50H. In the last three Gyr, 
the cluster can reach a large size ($\gtrsim10\,\pc$), but only near apocentre, where the largest size  ($\gtrsim20\,\pc$) is found for the cluster with a low background. In Fig.~\ref{fig:sdp_9403} we show the best fit number density profile for the model 50e50H (see Table~\ref{tab:UFO_param}). The flattening in the external region occurs where the number density of the model roughly equals to $\nBG$. 
 The simulated cluster is observed along the orbit, near apocentre, where the tails
overlap, in projection, with the cluster itself. From this we see that even though $\reff$ is in the correct size range, the Plummer profile
does not fit the data properly, because the stars in the tails
generate a bump in the number density profile, which is not observed for UFOs \citep{2008ApJ...684.1075M}. 
Whereas, near pericentre, where the tails are elongated, the number density profile is well reproduced by a Plummer model, see Fig.~\ref{fig:sdp_9103}. 

\subsection{The effect of the line-of-sight on the size measurements}
\begin{figure}
\center
\includegraphics[width=0.48\textwidth]{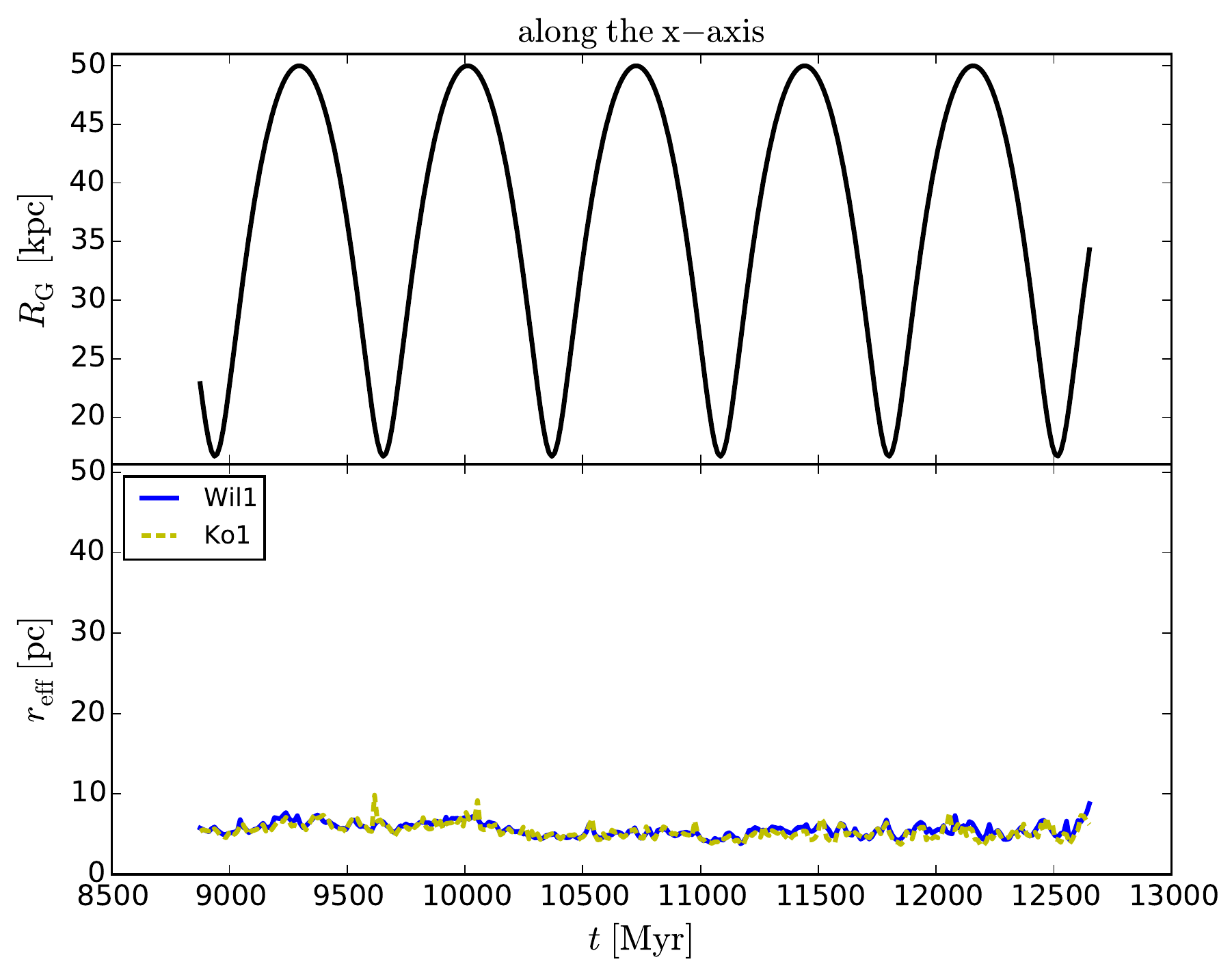}
\caption{
 Top: orbit of the cluster.
Bottom: Evolution of the half-number radius of the 50e50H model along the
    $x$-axis. As in Fig.~\ref{fig:comparex}, half-number radius for a cluster with
    a Wil1-like background (blue line) and a Ko1-like background (dashed yellow line).
}\label{fig:comparez}
\end{figure}

Although we showed that orbital phase can greatly influence the apparent size of
a cluster, an additional fact must be considered, which is the dependence on the
line-of-sight. To study different lines-of-sight, we translated the $N$-body
coordinates from a non-rotating frame, to a frame  in which the cluster is
orbiting in the $x$-$y$ plane with positive angular momentum centred on the
Galaxy and with the $x$-axis increasing towards the cluster. As shown in
Fig.~\ref{fig:comparex}, along the $y$-axis near apocentre the cluster reaches 
$\reff\sim20\,\pc$, while when viewed along the $x$-axis and $z$-axis, Fig.~\ref{fig:comparez},
 we do not see any variations in the $\reff$ measurements linked to the orbital motion. As a
consequence, these star clusters appear as eUFO only when observed 
 along the $y$-axis and when they are near apocentre. For all the simulations, the
results for the $x$ and $z$ directions are identical, therefore, in the
following figures we will show only one of them.

Because in our chosen reference frame, the $y$-axis is not along the orbit in
between pericentre and apocentre, we also considered the cluster's properties
along the orbit at those positions, to see whether the projected tails can
influence the measured cluster's size. Near apocentre and pericentre we expect
to obtain the same results as when we observe the cluster along the $y$-axis,
because in pericentre the two lines-of-sight overlap. For the entire 
evolution of the cluster, we found that the
estimation of the size along the $y$-axis and along the orbit are comparable.

The $y$-axis is the only line-of-sight along which we can observe clusters with a large size, 
however it is also the least probable one; because these objects are in the halo of the Milky Way ($\RG\gtrsim 20\,\pc$). Therefore, unless they have their pericentre within the solar circle, it is impossible to observe them along the orbit.

To quantify the probability to observe an eUFO, we estimated the fraction of orbit ($\ft$) in which a cluster appears extended. Therefore, $\ft$ is the ratio between the time when a cluster appears extended and its orbital period. For the simulation 50e50H along the $y$-axis $\ft\sim0.08$, but if we take into account the fact that along the other lines-of-sight $\ft=0$ then the probability to observe the cluster is $<1\%$.

\subsection{The effect of the initial cluster density on the size measurement}
An additional parameter to take into account is the initial cluster density. As
shown in Table~\ref{tab:UFO_param}, we divide our simulations in high density
(H) and low density (L) clusters.  The low density clusters fill the
Roche-volume initially while the high density clusters are initially
Roche-underfilling.

\citet{H61} showed that a cluster, once it has filled its Roche-volume, evolves
with a constant  ratio of $\rh$ over $\rJ$  (Sec.~\ref{sec:estimation size}). 
 Therefore, we expect that the evolutions of $\reff$ of the
clusters on the same orbit but with different initial densities are similar in
the final stage of evolution.

\begin{figure}
\center
\includegraphics[width=0.48\textwidth]{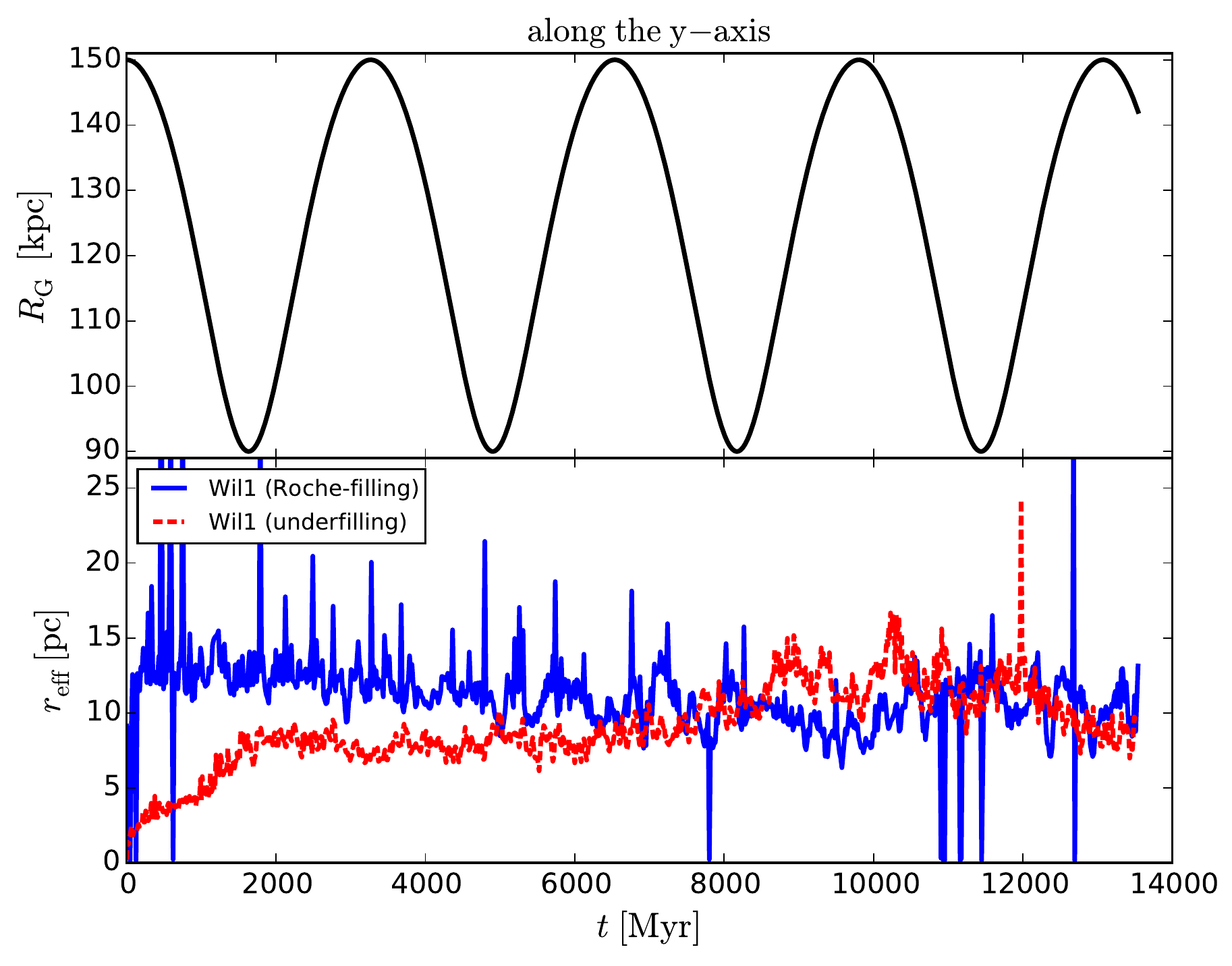}
\caption{
Top: orbit of the cluster.
Bottom: Evolution of the half-number radius of the 150e25H (dashed red line) and 150e25L
 (blue line) models along the $y$-axis.
 }\label{fig:comparex150e25}
\end{figure}

We find that this is indeed the case for most of our models
(Fig.~\ref{fig:comparex150e25}).  However, we find that there is a difference in
the evolution of $\reff$ depending on the initial density for three of our
orbits: $\RG=50\,\kpc$ and $\epsilon=0.75$; $\RG=50\,\kpc$ and $\epsilon=0.50$;
$\RG=100\,\kpc$ and $\epsilon=0.75$.

\begin{figure}
\center
\includegraphics[width=0.48\textwidth]{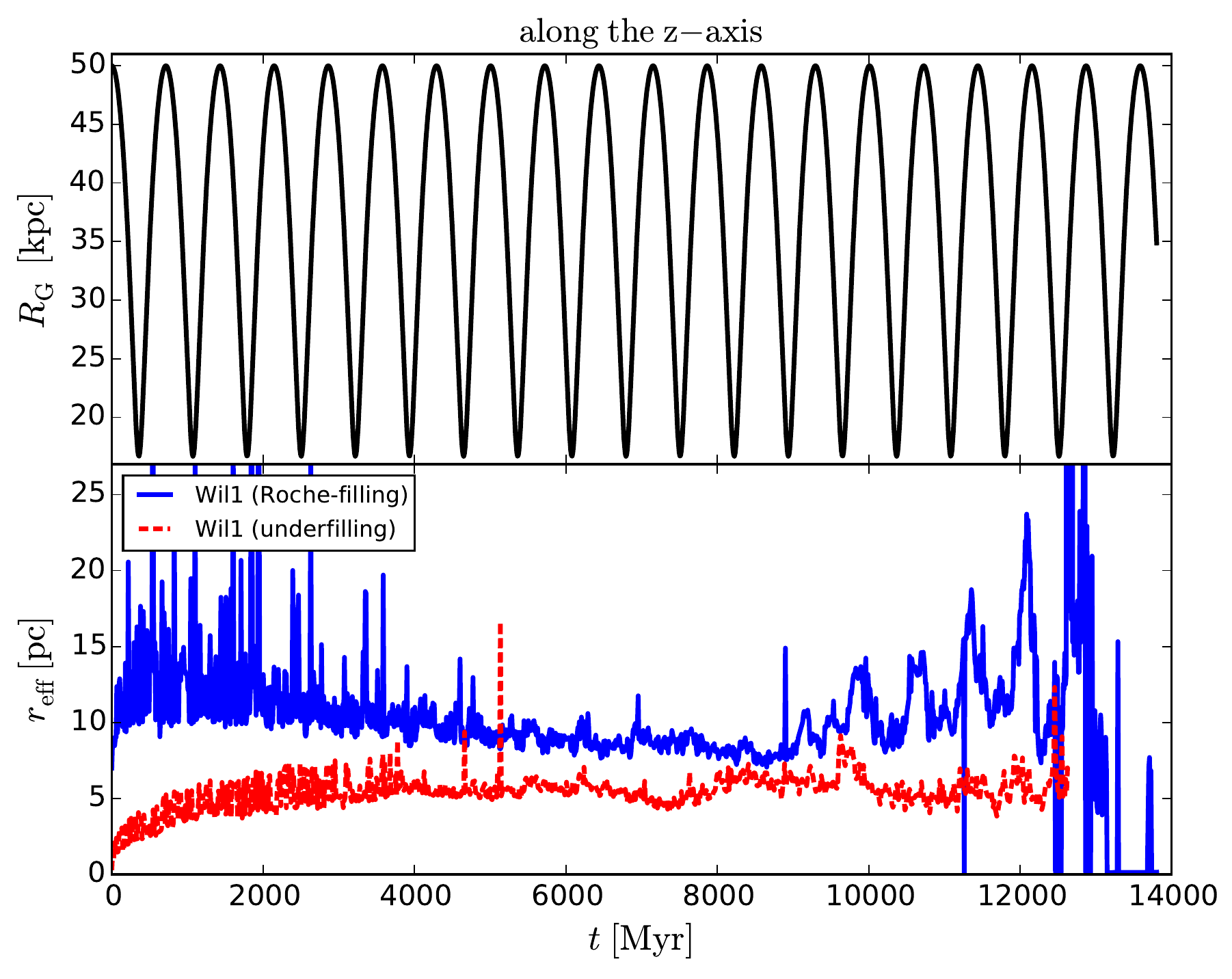}
\caption{
Top: orbit of the cluster.
Bottom: Evolution of the half-number radius of 50e50H (dashed red line) and 50e50L (blue line) models along the
$z$-axis (same results for the $x$-axis). Both the models have a
Wil1-like background.
}\label{fig:compareyLH}
\end{figure}

In Fig.~\ref{fig:compareyLH} we show an example of the $\reff$ evolution for two
models on the same orbit with different initial densities and it can be seen
that $\reff$ of the low-density cluster always lays above $\reff$ of the
high-density cluster. We interpret this difference as being due to the slow removal of stars in the
early evolution of the clusters with low densities that stay near the cluster
and can enhance $\reff$ at later stage. The high density cluster loses stars in
all directions with higher velocity in the initial phases, and these stars are
then too far to affect the $\reff$ measurement.  Furthermore, we observe a
greater variation of $\reff$ due to the orbital motion, visible in all the
lines-of-sight for these three orbits, in the simulations with a low initial
density. Whereas the clusters  with a high initial density appear larger only
along the $y$-axis. Therefore, to observe an extended cluster along all the lines-of-sight, this has to initially have a low density. For these simulations $\ft$ can be as high as $\sim0.54$, this estimate changes for different orbits and whether the cluster is close to dissolution.

\subsection{The effect of cluster's orbit on the size measurement}
\label{sec:cluster_orbit}
\begin{figure}
\center
\includegraphics[width=0.48\textwidth]{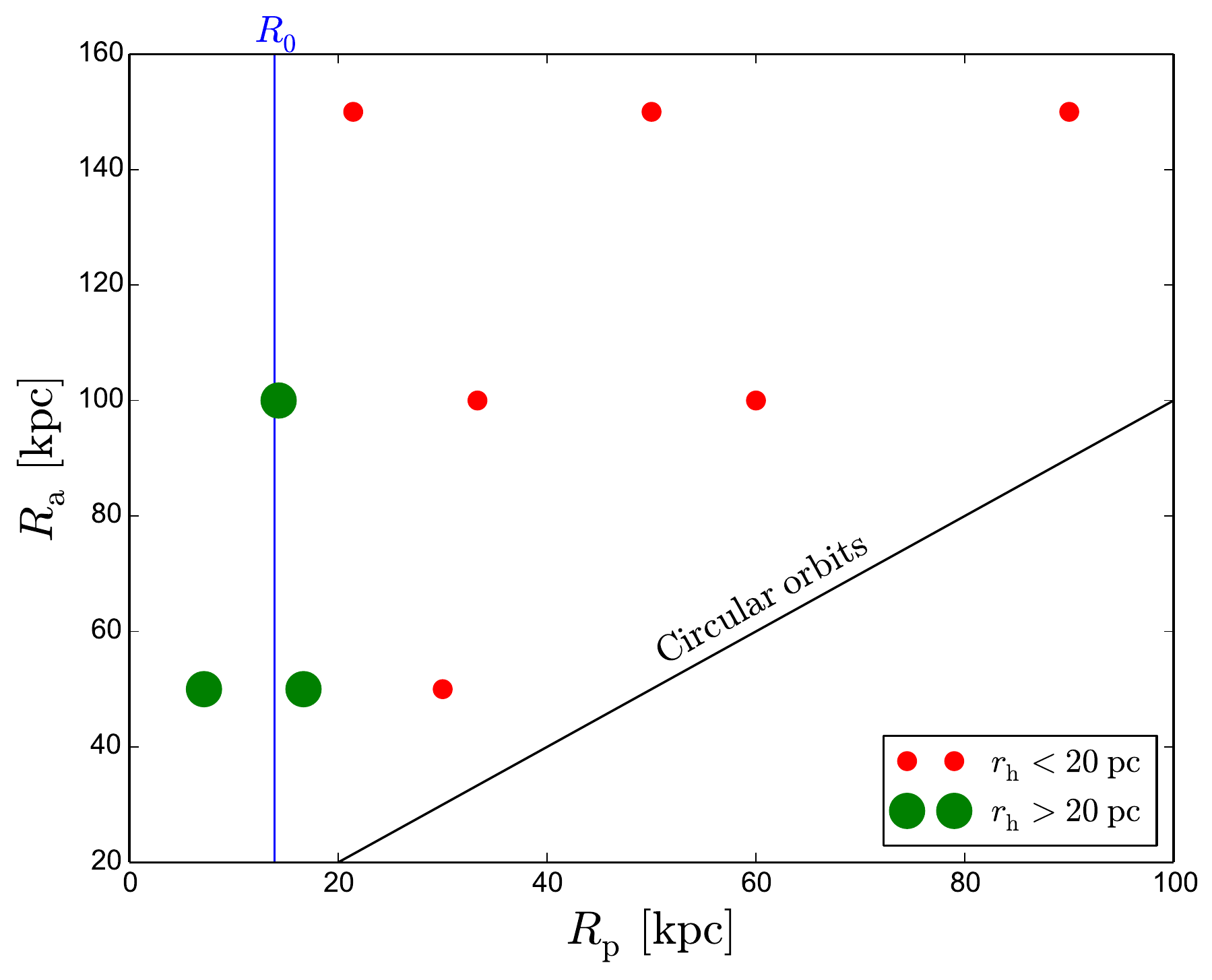}
\caption{
Effect of cluster orbits. The green dots show when we observe a cluster larger
than $20\,\pc$ in any line-of-sight, while the red dots show orbits of clusters that are always
smaller than $20\,\pc$. $R_0$ is the scale radius of the NFW potential. As
shown, the simulations that have their pericentre close or within the scale
radius reach a larger size. For initially Roche-filling clusters and close to dissolution, $f_t$ can be as high as $\sim0.19$, $\sim0.45$ and $\sim0.54$, for the models 50e50L, 50e75L and 100e75L, respectively.
}\label{fig:cluster_effect}
\end{figure}

To understand the relation between the orbit of a cluster and its $\reff$, we
illustrate in Fig.~\ref{fig:cluster_effect} the pericentre ($\Rp$) and the
apocentre ($\Ra$) of each orbit considered in this paper
 in a static NFW potential 
(Table~\ref{tab:UFO_param}). Therefore, each point represents an orbit.
The colours mark whether the size of a cluster can appear
larger than $20\,\pc$ (in green) or not (in red), due to the variation in the
size evolution, as shown in Fig.~\ref{fig:comparex}. 

Surprising, in Fig.~\ref{fig:cluster_effect} the three green dots represent 
the three orbits mentioned in the previous section (where $\reff$ evolution
for different initial density  
never overlap). 
We find that these three orbits have their $\Rp$ either close or within the
scale radius ($R_0$, blue vertical line) of the Galactic potential, where the
slope of the NFW density profile changes.  A variation of the Galactic density
profile implies a different evaporation mass loss of the cluster during its
pericentre passages. In these orbits, the stars that escape at $\Rp$ are easily
coming back close to the cluster, so that they can inflate the size measurement,
especially when the cluster is initially Roche-filling.

Among the simulations with a pericentre close to $R_0$, the simulations 150e75H
and 150e75L which have $\Rp=21.4\,\kpc$ do not appear larger. Therefore, we
assume that all the clusters with $\Rp<20\,\kpc$ appear larger.
Considering only orbits with pericentre within $20\,\kpc$ and the $\ft$ in the previous Section, the probability to observe an eUFO, that was initially Roche-filling, can be as high as $\sim30\%$.

\subsection{The effect of different potentials}
\label{sec:diff_potential}
To understand the role of the Galactic potential, we run additional simulations using different MW-like potentials; gNFW and P90.

In the previous section, we conclude that the scale radius of the Galactic
potential  has an important role to discern between star clusters and extended
star clusters. 

In a static potential, clusters orbiting around a galaxy have their pericentre
fixed in time (dynamical friction is negligible and has not been taken into account), while in a
growing potential, clusters that have their pericentre within the scale radius
at $12 \,\Gyr$ could have their pericentre beyond the scale radius initially.

Results from our simulations show that a gNFW potential 
(see Sec.~\ref{subsec:galactic_potential}) does not change the
evolution of star clusters, because the stars that inflate the size of a cluster
in the last few Gyr are the ones that have escaped recently from the cluster.  The properties of the simulations are presented in
Table~\ref{tab:UFO_param}.

\citet{2015arXiv150304815R} showed that cluster evolution does
not change in a gNFW potential also for satellites that have $\Ra\leq50\,\kpc$.
 In our case, we tested  clusters with $\Ra\geq50\,\kpc$, because, as described in \citet{2014A&A...563A.110B} 
 galaxies form inside out \citep{2003MNRAS.339..834H,2011MNRAS.413.1373W}. 
Which means that the mass of the MW is growing in shells by smooth accretion; 
therefore, the objects in the halo should be more affected by the 
growth of the DM potential.

After testing a growing halo potential, we studied the evolution of clusters in
a potential which includes a bulge, disc and halo component. The disc could
influence clusters which have their pericentre close to the Galactic centre.
Moreover, with a multi-component potential we can have non planar orbits,
increasing the probability to observe a cluster from different lines-of-sight.

To assess the effects of a multi-component potential, we run four simulations using a static potential for the bulge, the
disc and the halo, following the analytical model from P90 (see Sec.~\ref{subsec:galactic_potential}). 
The properties of the simulations are presented in Table~\ref{tab:UFO_param}. 

Even in this case the evolution of the observed size is similar to the
simulations with a NFW potential.
Therefore, we conclude that, with our initial conditions, 
the passage of a cluster through the disc does not enhance the size of the cluster, because the scale parameter of the disc ($\ad = 3.7\,\kpc$) is roughly half of the minimum pericentre distance($\Rp=7.14\,\kpc$).

\subsection{The effect of stellar mass black holes retained in the cluster}
\label{sec:BH}
There are no observational constraints on black hole (BH) natal kicks, 
while there are on neutron star natal kicks, mainly thanks to radio pulsars. 
For this reason, it is not clear whether 
the BHs natal kick should be similar \citep{2012MNRAS.425.2799R} 
or smaller than the neutron stars natal kick \citep{2012ApJ...749...91F}.
Likely, with the discovery of new gravitational waves, further
 constraints will be set on the BHs natal kick velocity \citep{2016ApJ...818L..22A}.

It has been shown by \citet{2004ApJ...608L..25M,2007MNRAS.379L..40M,2008MNRAS.386...65M,2013A&A...558A.117L} and \citet{2016MNRAS.462.2333P} that a higher fraction of dark remnants in a cluster can change its observed properties.
Moreover, BH candidates have been observed in several GCs \citep{2012Natur.490...71S,2013ApJ...777...69C}; as a consequence, we consider the possibility that BHs do not receive a kick 
 when they form and for these models we retain 100\% of stellar mass BHs initially.
The properties of the simulations are in Table~\ref{tab:UFO_param} and for the Galactic potential we assumed a NFW potential (eq.~\ref{eq:NFW}).

\begin{figure}
\center
\includegraphics[width=0.48\textwidth]{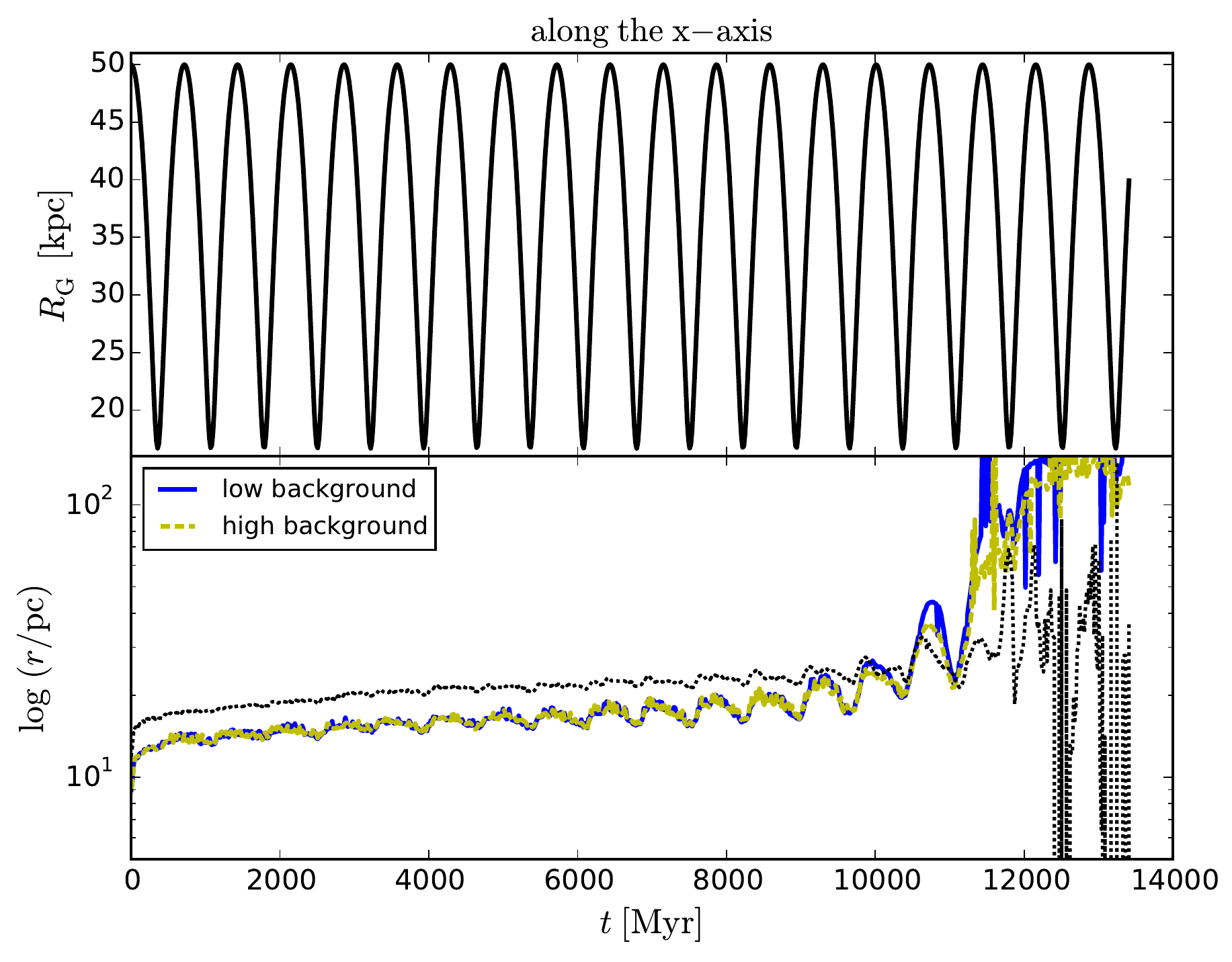}
\caption{In the lower plot, evolution of the half-number radius of the 50e50L-BH model along the $x$-axis. Half-number radius for a cluster with a Wil1-like background (blue line) and a Ko1-like background (dashed yellow line), 3D half-mass radius (dotted black line). In the upper plot the black line shows the orbit of the cluster.
}\label{fig:comparezBH}
\end{figure}

In Fig.~\ref{fig:comparezBH} we show the evolution of $\reff$ for the model 50e50L-BH that started with a low initial density. The clusters appear extended ($\reff \gtrsim 20\,\pc$) for almost the entire evolution 
(after roughly $9\,\Gyr$), independent of the projection axis 
(similar results for other lines-of-sight) and orbital phase. Indeed, the projection effect of the tails are not affecting the fitting results as in Fig.~\ref{fig:sdp_9403}.
Therefore, unlike the models that do not retain BHs, these clusters can be observed as eUFO ($\ft=1$). 

\begin{figure}
\center
\includegraphics[width=0.48\textwidth]{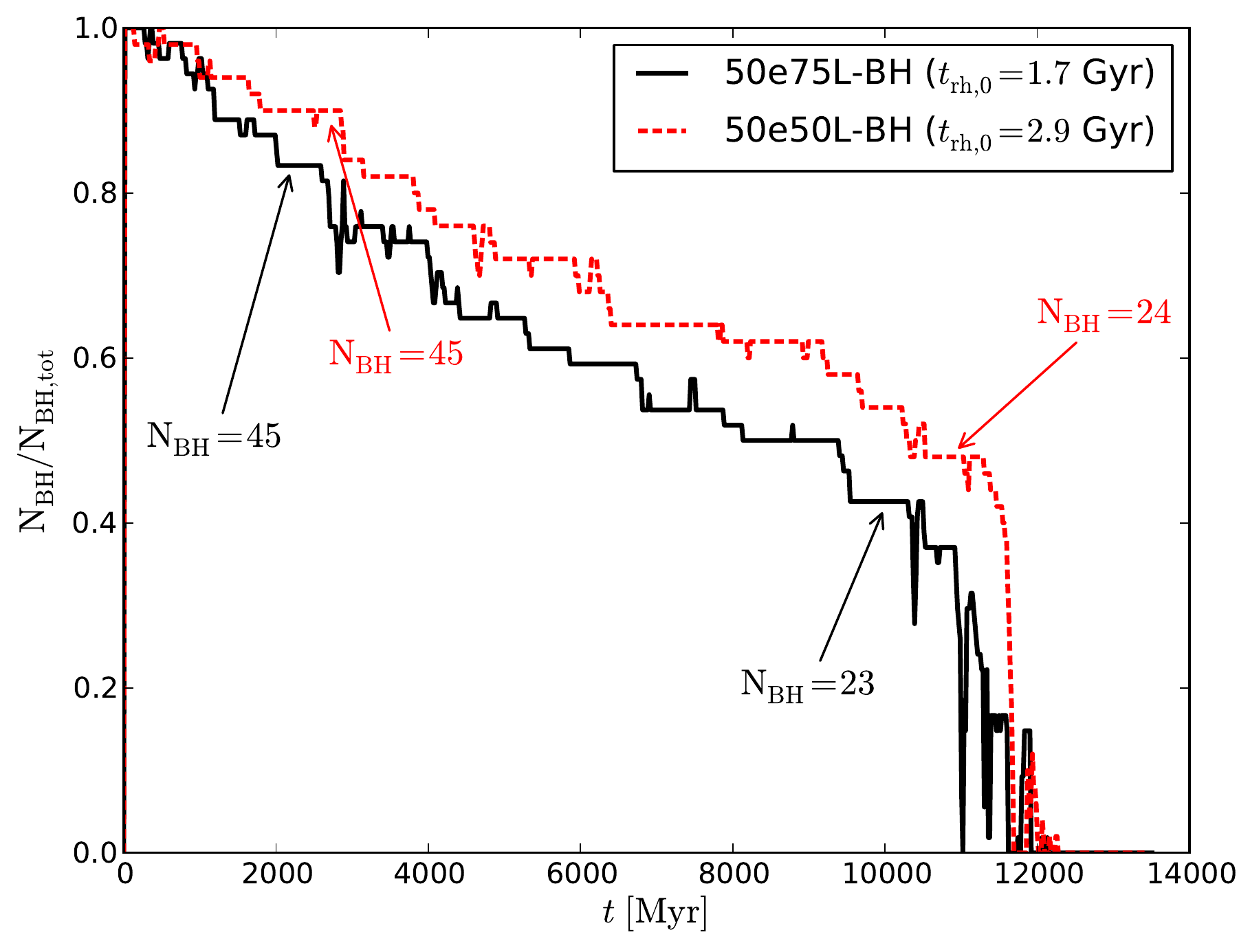}
\caption{Evolution of the fraction of BHs (normalized to the maximum value) 
for the models 50e50L-BH (dashed red line) and 50e75L-BH (black line), within the tidal radius of the clusters
}\label{fig:BH_BH0}
\end{figure}

In Fig.~\ref{fig:BH_BH0}, the evolution of the fraction of BHs inside the cluster (within the tidal radius) shows how fast the BH population escape from the cluster. 
\citet{2013MNRAS.432.2779B,2013MNRAS.436..584B} showed that the escape rate of stellar BHs depend on their half-mass relaxation time. 

For clusters with high initial density, for example in the simulation 50e50H-BH, because the short initial half-mass relaxation time ($\trhi$), all the BHs are dynamically ejected in few Gyr; indeed, the results are similar to the simulation 50e50H where only few percent of BHs are retained in the cluster initially.
Whereas, the low density clusters, which have a $\trhi$ of $\sim2-3\,\Gyr$, as shown in Fig.~\ref{fig:BH_BH0} they retain the BHs up to the dissolution of the cluster. 
These low density clusters do not appear mass segregated \citep{2016MNRAS.462.2333P}. 
The effect of stellar mass BHs retained in low density clusters is remarkable, because these clusters can appear as large as an eUFO for the last Gyr (not only near apocentre) and along all the lines-of-sight. However, in the absence of kinematics, it is challenging to determine whether these objects are DM free or dominated, because they do not appear mass segregated. Regarding their morphology, if we observe them along the $x$-axis (the most probable line-of-sight), we do not see the typical `S' shape of a star cluster, because the Lagrangian points (L1 and L2) overlap with the centre of the cluster. 

\subsection{Mass function}
\label{sec:MF}

 In this section we want to study the mass function (MF) of collisional system with
  large $\trhi$ ($\sim2-3\,\Gyr$), where the BHs are retained.

\begin{figure}
\center
\includegraphics[width=0.48\textwidth]{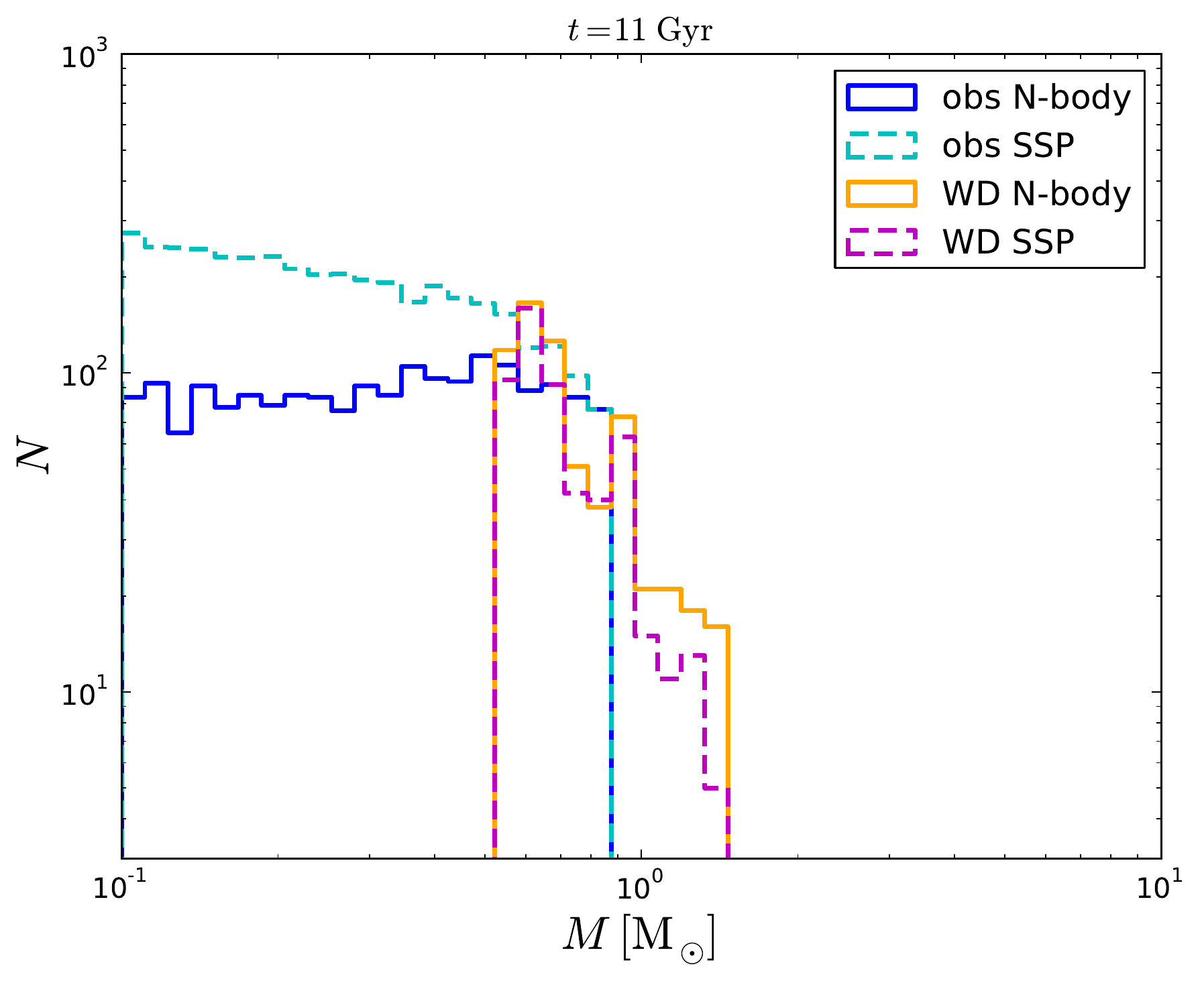}
\caption{Stellar mass function for the model 50e50L-BH and for a stellar population 
with the same IMF and age (SSP), but that has undergone no dynamical evolution, after $11\,\Gyr$.
}\label{fig:imf_bh}
\end{figure}

In Fig.~\ref{fig:imf_bh}, we plot the MF of the \nbody model 50e50L-BH, for all the stars without
 the dark remnants (blue line) and for the white dwarfs (WDs, orange line). Then we compare the MF
 of our model with a single stellar population (SSP) model. For the SSP model we assumed a Kroupa IMF
\citep{2001MNRAS.322..231K} between $0.1\,\msun$ and $100\,\msun$, and a  metallicity of $Z=0.0008$
(corresponding to $\feh\simeq-1.5$). We evolve the SSP model, up to $11\,\Gyr$, with  
the single-star evolution (SSE,
 \citealt{2000MNRAS.315..543H}) code,
  which is the same evolutionary tool available in {\sc nbody6}.
The MF of the SSP model is in dashed cyan line for all the stars except the dark remnants, 
and in dashed magenta line we plot the WDs. To compare these models we 
scaled the SSP MF to the \nbody MF, such that the number of stars in the last bin of the observable stars ($0.79<m/\msun<0.87$) is the same for SSP and \nbody model. From this comparison (Fig.~\ref{fig:imf_bh}), we can say that in collisional systems, where 
dynamical interactions between stars are important, the MF is flattened. 
Because ultra-faint dwarf galaxies appear to have similar MF slopes \citep[$\sim -1.3$ in the range $0.5-0.77\,\msun$,][]{2013ApJ...771...29G}, the flattened MF as a result of dynamical evolution can not be used to discern between extended star clusters and DGs for an individual object. However, for star clusters we do not expect a relation between the MF slope and the metallicity (as found for DGs), but we do expect the slope to be flatter at smaller $\RG$ \citep[e.g.][]{1997MNRAS.289..898V}. Hence the MF slope might be useful for addressing the nature of UFOs by considering the MF slope as a function of $\RG$ and $\feh$, simulteneously. 

As shown in Fig.~\ref{fig:imf_bh}, the model 50e50L-BH shows a large fraction of WD. To estimate how many WD are present in the models with respect to the observable stars, we estimate the number of WD ($\NWD$) between the first bin of the WD and the last bin of the observable stars, and the same for the number of observable stars ($\N*$).
Therefore, for $0.52<m/\msun<0.87$, $\NWD/\N* = 0.76$ for the SSP model and $\NWD/\N* = 1.12$ for the \nbody model. \citet{2012ApJ...760...78G} and \citet{2015ApJ...804...53H} show that, in the UV, the young WD are 
among the brightest stars in the cluster, which means
 that for FSC the WD population can potentially be observed. However, in a low-$N$ system such as a UFO, 
 the number of young WD is small.
 For example, with HST in the F225W band, for the model 50e50L-BH at $11\,\Gyr$ we expect to be able to observe only 4 out of 499 WD.

\subsection{Velocity dispersion}

To establish the dynamical mass of a system we need $\reff$ and the
velocity
dispersion, $\sigma$.  Despite the fact that it is challenging to determine
$\sigma$ for most UFOs, because of their distance and the limited number of bright
stars, for some of them the velocity dispersion has been
measured. For example, Wil1 has $\sigma\sim 0 \, \kms$ within $\reff$
\citep{2011AJ....142..128W}, which is consistent with a star cluster scenario;
whereas Segue 1 has $\sigma\sim 3.7$\footnote{Giant stars show a lower velocity dispersion, $\sigma\sim2^{+3.1}_{-1.7}\,\kms$.} $\,\kms$ within $\sim3\times\reff$ with a $V$-band mass-to-light 
ratio of $3400 \,\Msun/\Lsun$ 
\citep{2011ApJ...733...46S}, which lead to the conclusion that Segue 1 is a dark
matter dominated object.  We analysed our simulations keeping in mind the
observational biases discussed previously in this paper, and studied whether it
is possible to infer a high velocity dispersion in a dark matter free object. We
assumed that with a velocity measurement, member stars and background stars can
be separated,   and we therefore ignore the effect of background stars on the
$\sigma$ measurements.  Moreover, we studied the kinematics of the FSCs, to see
whether the orbital motion of the cluster could leave some features in the
velocity dispersion profile like in $\reff$ along the orbit near apocentre
(Fig.~\ref{fig:comparex}), as this has been the proposed scenario for the high
$\sigma$ of dwarf spheroidal galaxies (\citealt{1997NewA....2..139K} but see
\citealt{1997ASPC..116..259M} and \citealt{1998ASPC..136...70O}).  To compute
the $\sigma$, we are taking all the observable stars into account 
(see Sec.~\ref{sec:bg_effect}), within $\reff$, along a line-of-sight, 
as a function of time. 

Therefore, the velocity dispersion $\sigj$ is defined as: 

\begin{equation}
\sigj^2 = \frac{1}{(\Nh-1)}\sum^{\Nh}_i\left(v_{j,i}-\overline{v_j}\right)^2
\end{equation}
where $j$ is a chosen line-of-sight, $\Nh$ is the number of stars within the
projected half-number radius, $v_{j,i}$ is the velocity of the $i$-th star in
the line-of-sight and $\overline{v_j}$ is the mean line-of-sight velocity; and
we estimate the variance for each velocity dispersion as
\citep{1993ASPC...50..357P}

\begin{equation}
\Delta\sigj^2 = \frac{\sigj^2}{2{\Nh}}.
\end{equation}

\begin{figure}
\center
\includegraphics[width=0.48\textwidth]{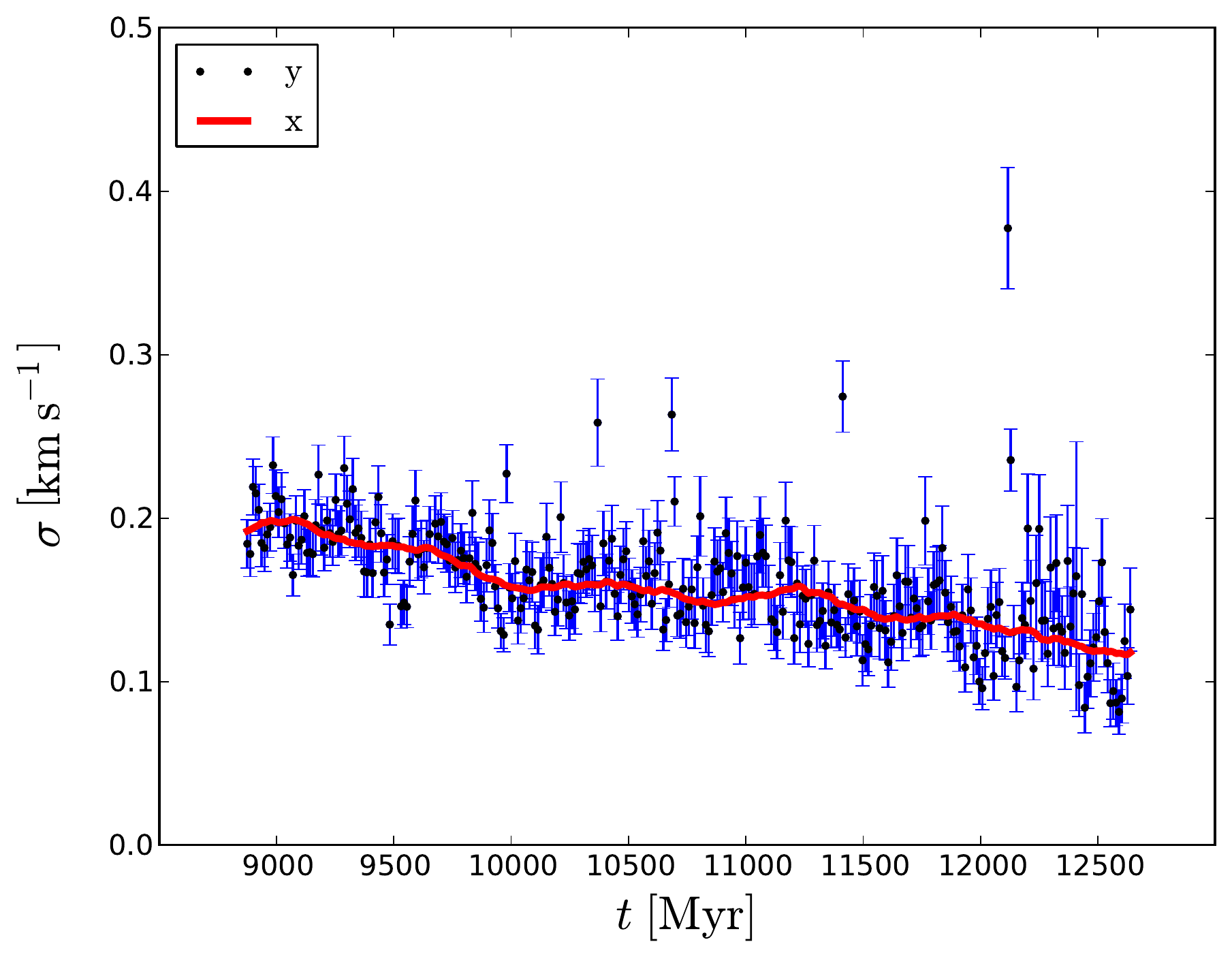}
\caption{
Evolution of the velocity dispersion of the model 50e50H along the $x$-axis (red line) and along the $y$-axis (black dot).
}\label{fig:sig}
\end{figure}

In Fig.~\ref{fig:sig}, we show the velocity dispersion along the $x$-axis (red
line) and along the $y$-axis (black dots). The other line-of-sight, $z$, is not
shown because it has the same trend and values of the $x$-axis.  As shown, there
are only features due to the orbital motion along the $y$-axis when the cluster
is near apocentre.  The cause of increase is similar to what we found for the
enhancement of $\reff$ along the $y$-axis, namely an increased number of unbound
stars projected within $\reff$.  Nevertheless, as shown in the
Fig.~\ref{fig:sig}, this rise happens only for a brief moment with respect to
the orbital period, near apocentre.  Because of this, and the fact
that it is impossible to view a system exactly along its orbit when it is in
apocentre (if the apocentre distance is further away than the solar radius),
this effect is not expected to play an important role in inflating the velocity
dispersion, at least not in the cases studied here. Therefore, any observation
of the velocity dispersion of a FSC without binaries in the outer halo should find
a value that is consistent with the virial mass of the stars and stellar
remnants (i.e. a few $100\,{\rm m}\,{\rm s}^{-1}$).

\subsection{Binaries}
\label{sec:Binaries}

Around one-third of the stars in the solar neighbourhood are in binaries or
multiple systems  \citep{2006ApJ...640L..63L} and UFOs may have a higher fraction of binaries
(e. g. \citealt{2011ApJ...733...46S,2011ApJ...738...55M}). Previous studies show that binaries play an important role in the cluster's evolution \citep{2003gmbp.book.....H,2005MNRAS.358..572I,2007ApJ...665..707H}. 
{\sc nbody6}  includes a prescription for both single star and binary star evolution
\citep{2000MNRAS.315..543H,2002MNRAS.329..897H} and allows us to study these
effects combined with their dynamical influence on the evolution of the cluster.
In this Section we focus our efforts in understanding and quantifying the effect
of primordial binaries on the velocity dispersion, performing three
simulations with $\sim20\%$ of primordial binaries (50e50M-B1, 50e50M-B2 and 50e50L-B2-BH). 

We report in Table~\ref{tab:UFO_param} the three runs. 
For 50e50M-B1 and 50e50M-B2 we have the same initial conditions as 
for the other clusters, except that the
initial density is lower with respect to the high density simulations. In these
two simulations, as for 50e50M, we have $\rhoh ={10^3 \, \msun \, \pc^{-3}}$ with an apogalacticon
of $50\,\kpc$ and eccentricity of 0.50. 

\begin{figure}
\center
\includegraphics[width=0.48\textwidth]{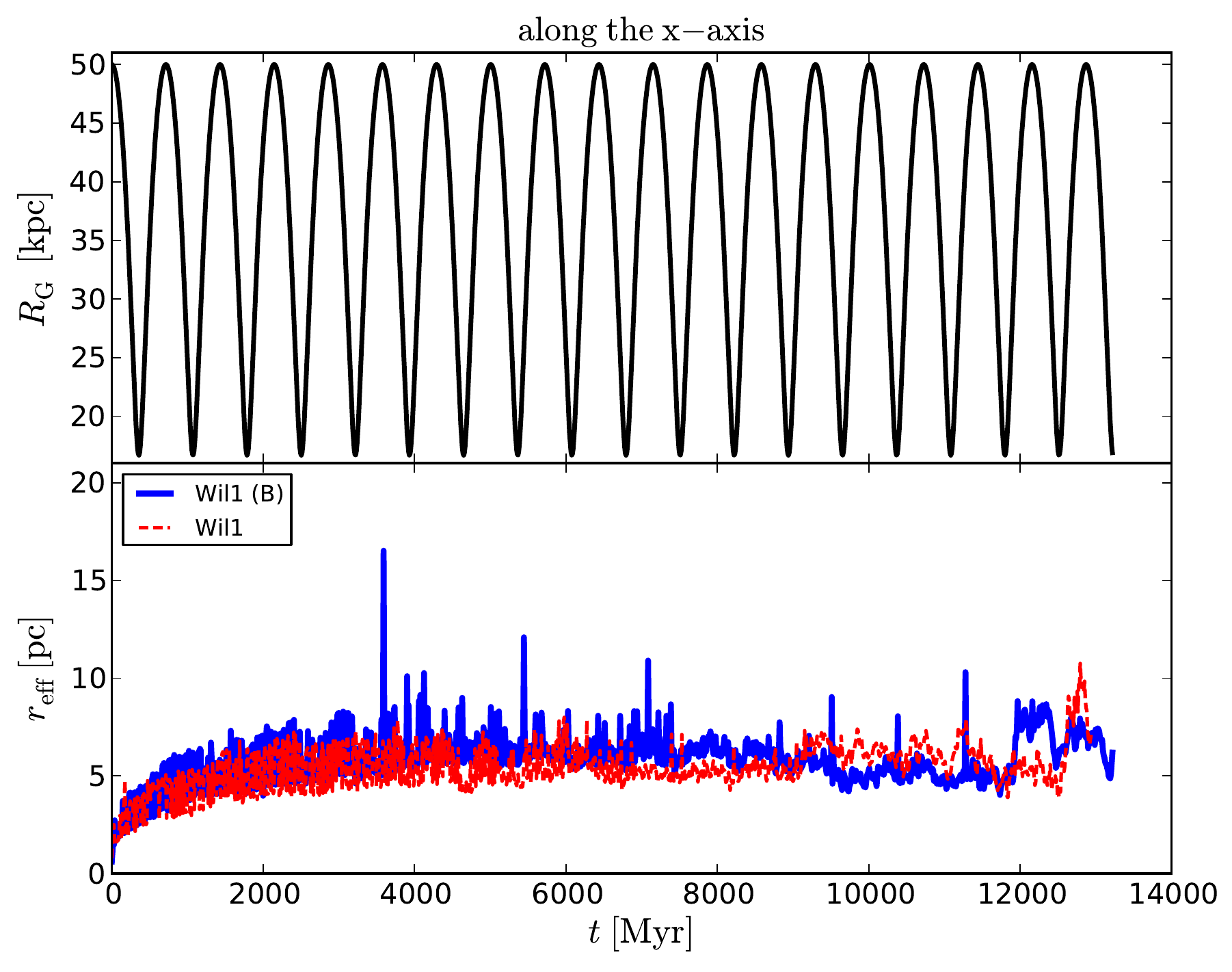}
\caption{
    Evolution of $\reff$ with (blue line) and without (dashed red line) primordial binaries along
    the $x$-axis. In blue $\reff$ of  50e50M-B1 and in red $\reff$ of 50e50M.
    The black line shows the radial orbit of the cluster.
}\label{fig:comparez09}
\end{figure}

In Fig.~\ref{fig:comparez09} we show the evolution of $\reff$ for the model with
binaries (50e50M-B1) and without (50e50M). The values for $\reff$ between the
two simulation are similar therefore we can conclude that primordial binaries do
not inflate $\reff$. Previous studies \citep{2011MNRAS.410.2698G} showed 
that the evolution of $\rh$ is insensitive to the binary fraction. 

For the analyses we treat stars in binary systems in the same way as the single
stars, i.e. if their luminosity is above the detection limit, we include them in
the analyses of $\sigma$. This means that there is an additional contribution to
$\sigma$ due to the orbital motion of the binary members.  As is often done in
observations, we apply a $\sigma$-clipping technique iteratively, removing all
the stars with velocities larger/smaller than  $3\sigma$ from the mean, until
the value of the $\sigma$ does converge \citep{1977ApJ...214..347Y}.

\begin{figure}
\center
\includegraphics[width=0.48\textwidth]{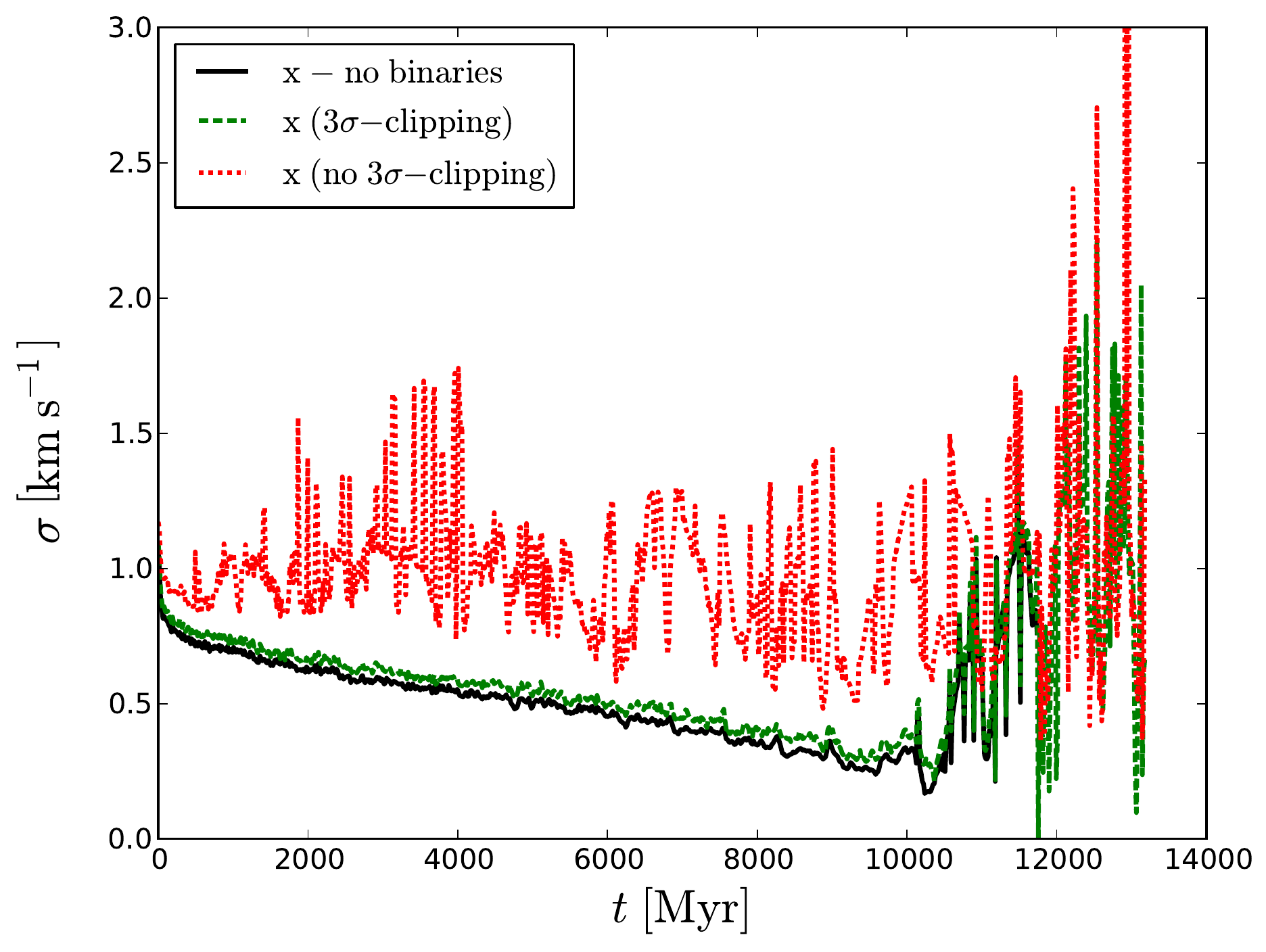}
\caption{Evolution of the velocity dispersion along the $x$-axis for the model 50e50L-B2-BH including all the binaries (dotted red line), applying the \mbox{$3\sigma$-clipping} technique (dashed green line), and without binaries (black line).
}\label{fig:sig09}
\end{figure}

In Fig.~\ref{fig:sig09} we show the evolution of the velocity dispersion for the 50e50L-B2-BH model; 
which dissolve after $11\,\Gyr$.
The increase in the observed $\sigma$ due to
binaries is small (green line) with respect
to the same model without primordial binaries (black line). Towards the end of
the cluster evolution we observe an increase in $\sigma$ associated with the
increased number of binary systems, this increase is due to preferential loss of
low-mass single stars \citep{2005MNRAS.358..572I,2007ApJ...665..707H}.
However, if the $\sigma$-clipping technique it is not taken into account (red line in Fig.~\ref{fig:sig09}), 
for example for a low number of observable stars, then the velocity dispersion
 is roughly $1\,\kms$. 
During the evolution of the cluster, the binary properties hardly change because the model has a large $\trhi$ (see Sec.~\ref{sec:BH}), while there typically only a few dynamically formed binaries (which have a short orbital period). Because the dynamical velocity dispersion is low ($\sim0.1\,\kms$), binaries with orbital velocities of $\sim0.5\,\kms$ are significantly affecting the inferred velocity dispersion. For a primary of $0.7\,\msun$ and a secondary $0.4\,\msun$  this corresponds to a period of $\sim1000\,{\rm yr}$, making it very challenging to detect these binaries in repeat observations. Because the binary properties do not evolve much, the only way of taking binaries into account would be to make an assumption about the binary properties and include this in the modelling \citep[e.g.][]{2011ApJ...738...55M, 2014A&A...562A..20C}. 

Assuming that the model \mbox{50e50L-B2-BH} at $10\,\Gyr$
  is in dynamical equilibrium,
 with the formula by \citet{2009ApJ...704.1274W} and \citet{2010MNRAS.406.1220W} we can estimate its 
 dynamical mass within the half-light radius of the system:
 \begin{equation}
 M_{1/2} = \frac{4}{G}\,\sigma^2 \,\reff \, .
 \end{equation}\label{eq:mdyn_estimate}
 
For example, for $\sigma \simeq 1\,\kms$ and $\reff\simeq30\,\pc$, $M_{1/2}\simeq2.8\times10^4\,\msun$.
From the simulation we can estimate the half-light luminosity $L_V \simeq 700\,\lsun$,
  therefore the $\ml \simeq 40 \,\mlsun$, which is consistent with a DM-dominated object interpretation.
  While if we consider the $3\sigma$-clipping, $\sigma \simeq 0.4\,\kms$ then $M_{1/2}\simeq4.5\times10^3\,\msun$ and $\ml \simeq 6 \,\mlsun$. Because $\reff$ and $\sigma$ change with time, in the cluster lifetime we have different estimates of the $\ml$, which can be as high as $100\,\mlsun$.

  Our estimation can be interpreted as a lower limit, because 
  we do not have background stars that can contaminate the measurements of the velocity dispersion.

\section{Conclusions}\label{sec:conclusion}
In this study we present \nbody simulations of the evolution of faint star clusters (FSCs)
 using {\sc nbody6tt}, which is an adaptation of the direct \nbody code {\sc nbody6}.
 We focus our analysis on the effects of​ observational ​biases on the 
 measurements of the properties of ultra faint objects (UFOs, see Fig.~\ref{fig:lumVSradius}).

To investigate whether UFOs are part of the (ultra-faint) DGs population, the GCs population or both; we first estimate the total number of FSCs based on a simple GC population synthesis model. This model assumes an initial distributions of star cluster masses and Galactocentric distances, which is then evolved by a simple mass loss prescription. We find that the predicted number of clusters ($N_{\rm FSC}=3.3^{+7.3}_{-1.6}$) is consistent with the number of observed star cluster candidates, see Fig.~\ref{fig:dNdMdRG}. However, more star clusters could be discovered and because we are using a very simple model, that for example ignores $\ml$ variations, we cannot conclude that part of the UFOs are DGs. 

Using a series of \nbody simulations we further study the biases that may arise from observing FSCs. These simulations were projected along different lines-of-sight and the object properties were derived using traditional state-of-the-art methods \citep{2008ApJ...684.1075M}. In order to investigate the role of the MW fore/background stars contamination we adopt a \trilegal-simulated field population at the position of two known UFOs (Koposov 1 and Willman 1). We find that the projected density of MW background stars can influence the measured size of a UFO in the sense that denser environments will lead to smaller sizes, when the Plummer model is not able to fit the number density profile properly (i.e. near apocentre and along the orbit, see Fig.~\ref{fig:sdp_9403}).

Besides the effect of the projected density of MW field stars, we notice that the observed size of a UFO depends on the orbital phase. Objects closer to apocentre tend to appear larger, however this effect is highly dependent on the viewing angle of the simulations, since the apparent larger size is caused by the overlapping of tidal tail stars. We find that the size enhancement is visible when the object is observed along the $y$-axis (the cluster is orbiting on the $x$-$y$ plane with positive angular momentum and the $x$-axis is the line that connect the Galactic centre to the cluster centre), for a small fraction of the orbit ($\sim8\%$, for the model 50e50H). 

The simulations also reveal a clear relation between cluster size and initial density in the sense that initially low-density clusters are more susceptible to size variations due to orbital phase (in all lines-of-sight). This result suggests that extended UFOs (eUFOs) are more likely to be observed if they formed Roche volume-filling. Furthermore, we observe that simulated clusters with pericentre roughly within the potential scale radius ($R_0$) show variations in the size measurements. This appear to be due to the change of the NFW density slope near the scale radius, which ultimately influences the strength of the tidal forces.

From our analysis we conclude that compact UFO satellites (e.g. Koposov 1 and 2) may naturally arise from a population of collisional systems.
However, it is very unlikely that star clusters, that do not initially retain stellar mass BHs, contribute to the eUFO population of a MW-like galaxy. The probability of observing an eUFO becomes insignificant if we consider that it has to be viewed along the $y$-axis (the least probable line-of-sight) and near apocentre. Nevertheless, if a cluster forms filling its Roche volume and has its pericentre within the scale radius, the probability to observe it rise up to 30\%. It means that among the FSC population, 1 over 3 could be observed extended. We find that these results hold even on multi-component (P90) or growing Galactic potential. 

We find that the retention of stellar mass BHs (no natal kicks) radically changes the results. When the cluster is initially Roche-filling, the observed size (as the 3-D half-mass radius) is expanding for its entire lifetime. In particular, after $9\,\Gyr$ the observed size grows above $20\,\pc$, independent of the line-of-sight and the fore/background. Whereas, when the cluster has an initial high density and retain 100\% of BHs, the BHs do not change the evolution of the size because they are rapidly ejected due to two-body interaction. This scenario is partially supported by observations of stellar mass BHs in several GCs \citep{2012Natur.490...71S,2013ApJ...777...69C}. Moreover, indirect evidence for BH in GCs comes from the large core radii \citep{2004ApJ...608L..25M,2007MNRAS.379L..40M,2008MNRAS.386...65M} and the absence of mass segregation in NGC6101 \citep{2016MNRAS.462.2333P}.

Finally, we study the effect of primordial binaries in our simulations, in particular their effect on the velocity dispersion. If we combine the size and the velocity dispersion measurements from the simulations with stellar mass BHs and primordial binaries, we estimate a $\ml\sim1-100$. High $\ml$ values ($\ml\gtrsim10$) are observed in DGs \citep{2012AJ....144....4M}, which are DM dominated objects, but our simulations show that these measurements are not conclusive for a DGs interpretation. We show that the binaries can inflate the velocity dispersion and that in an extended star cluster the properties of the binaries do not change significantly. Therefore, binaries with different initial properties may inflate the velocity dispersion and $\ml$ even more. For this reason, because the initial properties of the binary population is unknown we have to rely on assumptions which may not be correct. However, \citet{2016MNRAS.461L..72P} show that wide binaries can be used to constrain the central distribution of DM in ultra-faint DGs. In addition to that, in some of the \mbox{eUFOs}, metallicity spread has been observed, which is indicative of an extended star formation history. This leads to the conclusion that the satellite is either a tidally disrupted DM-free galaxy, or a DG.

DGs have lower or similar metallicities than GCs but they also have a relaxation time
 longer than the Hubble time, which means that the dynamical evolution due to two-body relaxation is not important. 
 Therefore, they have a mass function (MF) which is not depleted in low-mass stars as a result of dynamical evolution. However, \citet{2013ApJ...771...29G}  showed that the MF of DGs becomes flatter with decreasing metallicity, which they attribute to IMF variations. In GCs, mass segregation and evaporation can change the slope of the MF during the evolution \citep{1997MNRAS.289..898V,2003MNRAS.340..227B}, hence 
we expect GCs to be depleted in low-mass stars near the end of their lives, and to have a flatter MF for smaller Galactocentric distance.

eUFOs are likely to be accreted objects from DGs interacting with the MW, 
because if they form in a DG, the probability to be initially Roche-filling 
(low density) is enhanced (e.g. \citealt{2008ApJ...672.1006E}). Therefore, if they form
with a low density they have a large initial half-mass relaxation time and only few BHs will be expelled 
due to dynamical interactions. Moreover, an initially DM-dominated 
object will be likely to retain a high number of BHs even when natal kicks are taken into account. 
The BHs will sit in the centre pushing out the low mass particles due to two-body relaxation.
 Then, when the dissolution of the system occurs with few hundreds of stars left, a faint 
 DM-free object can be observed.

Our results to some extend agree with \citet{2016arXiv160608778D} results,
 where they claim that Segue 1, an eUFO, can be a DM-free object.
 Unfortunately, it is not trivial to compare our results with their results,
 because we are using a direct \nbody code, ideal for collisional systems,
  while they are using a particle-mesh code, which is not ideal 
  to simulate star clusters 
 but less time consuming, as they stated in their conclusion.
 Therefore, they do not have stars with different masses and binaries stars, 
 which in our cases are fundamental to increase the observed velocity dispersion.
 However, our simulations are not fine tuned for Segue 1. 

In this paper, we conclude that star clusters contribute to both the compact and the extended population of UFOs. Retaining stellar mass BHs in an initially low density cluster is vital to have extended star cluster. While a high binary fraction can inflate the velocity dispersion measurements significantly, leading to the conclusion that the object has a high $\ml$ ratio. It is possible to say something about the nature of star clusters that appear as an eUFO by considering the kinematics of the (tidal) tails, because in the case of dissolving star clusters these should be cold (few $100\,\ms$). Therefore, if the UFO is a star cluster we expect to observe a flatter MF for smaller Galactocentric distance and uncorrelated with metallicity,  and dynamically cold tails; while mass segregation and binary properties cannot be used to discern between DM free and DM dominated object.

\section*{Acknowledgments}\label{acknowledgments}
MG acknowledges financial support from the Royal Society in the
form of a University Research Fellowship (URF) and an equipment
grant used for the GPU cluster in Surrey.
All authors acknowledge support from the European Research Council (ERC-StG-335936, CLUSTERS).
The authors thank Oscar Agertz, Florent Renaud, Alice Zocchi, Vincent H\'enault-Brunet, Justin I. Read, Alessia Gualandris for interesting discussions and the referee for comments and suggestions. We are grateful to Sverre Aarseth and Keigo Nitadori for making {\sc nbody6} publicly available, and to Dan Foreman-Mackey
for providing the {\sc emcee} software and for maintaining the online documentation; we also thank Mr David Munro of the University of Surrey for hardware and software support.
The analyses done for this paper made use of {\sc scipy} \citep{scipy}, {\sc numpy} \citep{numpy}, and {\sc matplotlib} \citep{matplotlib}.

\bibliographystyle{mnras}

\appendix
\section{Absolute magnitude in \textit{V}-band}\label{app:MV}
To compute the $V$-band absolute magnitude ($\MV$) of the simulated clusters we applied two methods. 

1) Knowing the luminosity ($L$ in ${\rm L_\odot}$) and the temperature ($T$ in ${\rm K}$) 
of each star ({\sc nbody6} output) is possible to calculate $\MV$.
\begin{equation}\label{eq:MV_cluster}
\MV=-2.5\log{\sum_{i=1}^N 10^{-0.4\MVi}}
\end{equation}
here $\MVi$ is the absolute magnitude in band $V$ of the $i$-th star and $N$ is the total number of stars.
\begin{equation}
\MVi= M_{V,\odot} -2.5\log\left(\frac{L_i}{\rm L_\odot}\right) - {\rm BC}
\end{equation}
where $M_{V,\odot}=4.8$ is the absolute magnitude of the Sun and BC is the bolometric correction:
\begin{equation}
{\rm BC} = 2.324497 + 2.5 \log(g(T_i))
\end{equation}
with 
\begin{equation}
g(T_i) = BB(\lambda,T_i) \cdot \Delta_\lambda
\end{equation}
where $\Delta_\lambda=88\cdot10^{-9}\,$m is the full width at half maximum (FWHM) of $V$-band filter. While $BB(\lambda,T_i)$ is the normalised black body radiation formula (Planck law):
\begin{equation}
BB(\lambda,T_i) = \frac{1}{BB_{tot}(T_i)}\frac{BB_0(\lambda)}{e^{\frac{hc}{\lambda K_{\rm B} T_i}}-1}
\end{equation}
with 
$BB_0 = \frac{2hc^2}{\lambda^5}$ and $BB_{tot}=\frac{\sigma}{\pi}T_i^4$.
Where $c$ is the speed of light, $h$ is the Planck constant, $\sigma$ is the Stefan-Boltzmann constant, $K_{\rm B}$ is the Boltzmann constant.
In our case we use $\lambda=551\,$nm, which is the central wavelength for the $V$-band filter.

2) We compute the absolute magnitude in band $V$ using the initial mass of the stars in the simulations.

From eq.~(\ref{eq:MV_cluster}) we need to compute $\MVi$. Using CMD 2.7, we can use the PARSEC isochrones v1.2S \citep{2012MNRAS.427..127B,2014MNRAS.444.2525C,2014MNRAS.445.4287T}, where the initial mass ($m_i^{model}$) of the stars and their absolute magnitude in $V$-band ($\MVi{^{model}}$) are given for a selected time.

At this point, we can generate a function which interpolate these data, therefore, we have the absolute magnitude in $V$-band as a function of the initial mass, $\MVi{^{model}}(m_i^{model})$. Using the initial mass of the surviving stars at fixed time in our simulations, we can estimate their absolute magnitude in $V$-band, 
$\MVi{^{model}}(m_i^{N \textrm{-body}})$.

In conclusion the two methods are equivalent because we obtain similar results.

\end{document}